\documentclass[authoryear, preprint]{elsarticle}

\usepackage{blindtext}
\usepackage{geometry}
\usepackage{bm}
\usepackage{graphicx}
\usepackage{subfigure}
\usepackage{enumitem}
\usepackage{amssymb}
\usepackage[linesnumbered,ruled,vlined]{algorithm2e}
\usepackage{multirow}
\usepackage{placeins}
\usepackage{amsmath}
\usepackage{mathtools}
\usepackage{makecell}
\usepackage{amsthm}
\usepackage[colorlinks]{hyperref}
\usepackage[title]{appendix}
\usepackage{booktabs}
\usepackage{float}

\theoremstyle{definition}
\newtheorem{definition}{Definition}
\theoremstyle{remark}
\newtheorem*{remark}{Remark}
\newcommand\coolleftbrace[2]{%
#1\left\{\vphantom{\begin{matrix} #2 \end{matrix}}\right.}
\journal{arXiv}

\begin{document}

\begin{frontmatter}

\title{A Probabilistic Approach for Queue Length Estimation Using License Plate Recognition Data: Considering Overtaking in Multi-lane Scenarios}
\author[1]{Lyuzhou Luo}
\author[1]{Hao Wu}
\author[1]{Jiahao Liu}
\author[1]{Keshuang Tang\corref{cor}}
\ead{tang@tongji.edu.cn}
\author[2]{Chaopeng Tan\corref{cor}}
\ead{c.tan-2@tudelft.nl}
\cortext[cor]{Corresponding author}
\address[1]{Key Laboratory of Road and Traffic Engineering of the Ministry of Education, College of Transportation Engineering, Tongji University, Cao'an Road 4800, Shanghai 201804, China}
\address[2]{Department of Transport and Planning, Delft University of Technology, Gebouw 23, Stevinweg 1, 2628CN, Delft, Netherlands}

\begin{abstract}
Multi-section license plate recognition (LPR) data provides input-output information and sampled travel times of the investigated link, serving as an ideal data source for lane-based queue length estimation in recent studies. However, most of these studies assumed the strict first-in-first-out (FIFO) rule or a specific arrival process (e.g., uniform arrival), thus ignoring the potential impact of overtaking and the variation of traffic flows, especially in multi-lane scenarios. To address this issue, we propose a probabilistic approach to derive the stochastic queue length by constructing a conditional probability model of \textit{no-delay arrival time} (NAT), i.e., the arrival time of vehicles without experiencing any delay, based on multi-section LPR data. First, the NAT conditions for all vehicles are established based on upstream and downstream vehicle departure times and sequences. To reduce the computational dimensionality and complexity, a dynamic programming (DP)-based algorithm is developed for vehicle group partitioning based on potential interactions between vehicles. Then, the conditional probability of NATs of each vehicle group is derived and a Markov Chain Monte Carlo (MCMC) sampling method is employed for calculation. Subsequently, the stochastic queue profile and maximum queue length for each cycle can be derived based on the NATs of vehicles. Eventually, to leverage the LPR data sufficiently, we extend our approach to multi-lane scenarios, where the problem can be converted to a weighted general exact coverage problem and solved by a backtracking algorithm with heuristics. Empirical and simulation experiments have shown that the proposed approach outperforms the state-of-the-art method, demonstrating significant improvements in accuracy and robustness across various traffic conditions, including different V/C ratios, matching rates, and FIFO violation rates. In addition, the performance of the proposed approach can be further improved by utilizing multi-lane LPR data.
\end{abstract}


\end{frontmatter}

\section{Introduction}

Queue length is a crucial metric reflecting the supply-demand relationship and coordination quality at signalized intersections \citep{yao_dynamic_2020, noaeen_real-time_2021}. Accurate estimations of lane-based queue lengths could also significantly improve signal timings and active queue management at signalized intersections \citep{ma_back-pressure-based_2021, yin_queue_2021, tan2024connected}.

Early research on queue length estimation has depended on traffic detection data such as volume, occupancy, and speed collected by fixed-location detectors \citep{sharma_inputoutput_2007, vigos_real-time_2008, skabardonis_real-time_2008, liu_real-time_2009, wu_identification_2010}. But in reality, these methods often face limitations related to aggregation intervals, malfunctions, and high maintenance costs \citep{ban_real_2011}. With the development of new data sources like connected vehicles (CV), a considerable amount of real-time trajectory data is becoming available for traffic management at urban road networks \citep{tan2024privacy}. Compared with traditional fixed-location detectors, the connected vehicles can provide continuous position and motion information without any additional expense, thus were used for queue length estimation by deterministic \citep{cheng_exploratory_2012, ramezani_queue_2015, li_real-time_2017, tan_cumulative_2022} or stochastic \citep{comert_analytical_2011, zhang_cycle-based_2020, tan_cycle-based_2021} approaches. However, since CV trajectory data are sampled observations, these methods typically suffer from low penetration rates or infrequent data uploads.

Vehicle license plate recognition (LPR) data are another type of emerging data source on urban roads. LPR systems are installed near the stop line of signalized intersections and can collect real-time departure information on individual vehicles, including license plate numbers, vehicle types, lane, and appearance times before the stop line. Compared to aggregated data collected from loop detectors and sparsely sampled data from probe vehicles, the LPR data offers high quality, wide coverage, and full-sample detection. Furthermore, by matching license plates collected by LPR systems installed at adjacent intersections, individual vehicles' paths and travel times can be obtained. This feature has been exploited at the network level for OD flow estimation \citep{rao_origin-destination_2018, mo_estimating_2020, tang_dynamic_2021} and vehicle path reconstruction \citep{yang_vehicle_2015, cao_tracking_2024}. While vehicle entry and exit times can be directly obtained at the intersection or link level, vehicle behaviors within the link remain unknown. Thus, recent research has primarily focused on estimating traffic parameters not directly retrievable from LPR data, such as queue length \citep{wu_left-turn_2019, tang_lane-based_2022, tan_extendable_2022, zhan_link-based_2020, zhan_lane-based_2015, ma_estimating_2018, luo_queue_2019}, traffic demand \citep{ma_traffic_2017,an_lane-based_2021} and speed profile \citep{mo_speed_2017}.

Regarding queue length estimation using LPR data, existing methods can usually be separated into two broad categories: single-section methods and multi-section methods. Single-section methods solely rely on LPR data collected from target lanes and extract patterns from the vehicle discharging sequences for queue length estimation, e.g., critical point analysis (CPA) method \citep{wu_left-turn_2019}, E-Divisive with Medians (EDM) method \citep{tang_lane-based_2022}, Gaussian mixture model-based method \citep{tan_extendable_2022}, and Gaussian process model-based method \citep{zhan_link-based_2020}. Despite the minimal deployment requirements for LPR systems, the performance of single-section methods drops dramatically when traffic conditions become near-saturated or over-saturated. This limitation arises because single-section LPR systems can only capture link departure information, not arrival information, making these methods unsuitable for scenarios involving residual queues.

In contrast, multi-section methods employ LPR data from both the target lane and its upstream intersection. \citet{zhan_lane-based_2015} modeled the arrival flow rate by a piecewise linear function and developed a Gaussian process-based interpolation method to reconstruct each lane's equivalent cumulative arrival–departure curve. The arrival times for unmatched vehicles were obtained by presuming the validity of first-in-first-out (FIFO), and then the queue lengths were estimated in a car-following-based simulation scheme. Assuming uniform arrival rates, \citet{ma_estimating_2018} introduced two shockwave theory-based models focusing on the leading vehicle's queued position and the cycle maximum queue length and a recursive formula for maximum queue lengths across cycles, respectively. By combining these models, the potential relationship between maximum delay time and maximum queue length in each cycle was achieved. Without the consideration of overtaking, \citet{luo_queue_2019} focused on the intrinsic connections between the travel time of individual vehicles and the queue composition in each cycle, then the queue length in the immediate past cycle was formulated as a prediction problem using regression analysis. \citet{wu_stochastic_2024} proposed a stochastic queue profile estimation method, considering the scenario where all vehicles entering and exiting the road sections are detected by LPR systems. The arrival and departure curves were obtained using the input-output model, and the pseudo-departure curves were derived by considering the distribution of free-flow travel time. Then the stochastic queue profile was reconstructed by analyzing these three curves.

Current research using multi-section LPR data generally treats multi-lane scenarios by estimating queue lengths for each lane independently, without integrating data across lanes to achieve a comprehensive multi-lane estimation. This approach disregards information provided by upstream unmatched vehicles that may depart from other lanes downstream, thereby overlooking the interactions between lanes that are crucial for accurate queue length estimation. Moreover, these studies typically involve two strict assumptions that influence the estimation of queue lengths. The first assumption is the FIFO rule \citep{zhan_lane-based_2015, luo_queue_2019}, which implies that vehicles traveling towards a specific lane do not overtake each other, allowing for the direct extraction of arrival sequences from departure sequences. This assumption often fails in multi-lane scenarios due to frequent overtaking, leading to significant variances in predictions. The second assumption is a specific type of arrival process, either uniform \citep{ma_estimating_2018} or piecewise linear \citep{zhan_lane-based_2015}, which do not adapt well to the real and dynamic nature of traffic flows. Notably, \citet{wu_stochastic_2024} attempts to relax the FIFO assumption by considering overtaking in its model for queue length estimation based on LPR data, which involves enumerating various overtaking cases to create a pseudo-departure curve. However, this method requires fully sampled LPR data without any false detections and fails to account for hidden overtaking behaviors that are captured in the matched license plates (between the upstream intersection and the target lane).

In response to the identified limitations of current methods in queue length estimation using multi-section LPR data, we propose a novel probabilistic approach that incorporates overtaking behaviors and utilizes data from multiple lanes for enhanced performance. The proposed approach begins with the concept of a vehicle's \textit{no-delay arrival time} (NAT). We adopt this concept proposed by \citet{ban_real_2011}, which indicates the projected time a vehicle would have arrived at the intersection if no delay had been experienced. By establishing conditions that vehicles must satisfy in NATs and applying a dynamic programming (DP)-based partitioning algorithm, we efficiently categorize vehicles into constrained and unconstrained groups. Then, a Markov Chain Monte Carlo (MCMC) sampling algorithm is used to estimate the arrival distribution of each vehicle based on the prior running time distribution. After that, the stochastic queue profile and maximum queue length for each cycle can be obtained by accumulating arrival distributions of vehicles. Additionally, we extend this approach from the single-lane estimation to the multi-lane estimation by identifying an optimal matching between unmatched vehicles with a heuristic algorithm, which can reduce the uncertainty in queue length estimation and achieve better estimation results.

The contributions of this paper are summarized as follows:

\begin{enumerate}
\item We consider overtaking and its potential effects on other vehicles in queue length estimation based on LPR data, thereby relaxing two commonly used assumptions in existing studies: the FIFO rule and a specific arrival process. This relaxation allows our method to reflect real-world conditions better and apply to complex traffic scenarios.
\item Unlike existing studies, our approach can be extended to a comprehensive multi-lane estimation, rather than treating each lane separately. This allows us to fully exploit LPR data, including those unmatched vehicles that cannot be handled by existing studies.
\item Our proposed probabilistic approach not only enables high-resolution queue profile estimation but also provides confidence intervals. Evaluation results show that our proposed approach performs well in both empirical and simulation case studies, achieving higher accuracy than the state-of-the-art method \citep{zhan_lane-based_2015} in various data scenarios for both maximum queue length and queue profile estimation.
\end{enumerate}

\section{Terminologies and assumptions}\label{sec:term&assmp}
In this paper, we define the terminologies corresponding to the queue length as follows:

\begin{itemize}
    \item \textit{Queue length}: the number of vehicles required to span the distance between the last queued vehicle and the stop line. It remains unchanged until the last queued vehicle starts to depart, even when the queue begins to dissipate.
    \item \textit{Queue profile}: the queue lengths recorded at each time step, representing the profile of the queue formation wave. While other studies examine both queue formation and dissipation waves, our research focuses primarily on the queue formation process, as this is not directly accessible in LPR data.
    \item \textit{Maximum queue length}: the maximum queue length observed during a given cycle, from the start of the red light until the queue length begins to decrease. Note that, in cases of high saturation, the queue length may not decrease until the next cycle, but it is still considered the maximum queue length for the current cycle.
\end{itemize}

Other terminologies used in the methodology are summarized as follows:
\begin{itemize}
    \item \textit{Running time}: The total time required to traverse the link, excluding the stopped delay \citep{berry_distribution_1951}.
    \item \textit{No-delay arrival time} (NAT): the projected time a vehicle would have arrived at the intersection if no delay had been experienced. The values for each vehicle $k$ can be denoted as $t_k^a$. Unless stated otherwise, ``arrival distribution'' in the paper refers to the distribution of NATs, and ``NAT conditions'' refers to conditions that need to be satisfied for each vehicle's NAT within each group, based on actual observations.
    \item \textit{Matched vehicle}: vehicles captured by both upstream and the target lane and passed the license plate matching process.
    \item \textit{Unmatched vehicle}: vehicles filtered out in the license plate matching process.
    \item \textit{Constrained group}: For vehicles in a constrained group, both the first and the last vehicles must be matched, then the NATs for other vehicles within the group will be constrained between the two vehicles, including any unmatched vehicles. 
    \item \textit{Unconstrained group}: Unconstrained groups are separated by constrained groups, and vehicles within are all unmatched. Without any upstream information, the NATs of vehicles in an unconstrained group can be determined after calculating the arrival distribution of adjacent constrained groups. 
    \item \textit{Inter-group departure gap}: the minimum time difference between when vehicles depart from the upstream in adjacent constrained groups. An adequate departure gap is necessary to maintain low interaction between adjacent constrained groups.
\end{itemize}

The main assumptions used in the methodology are defined as follows:
\begin{enumerate}
    \item Miss detections of vehicles are not considered, as the main source of error in current LPR systems lies in false detections \citep{zhan_lane-based_2015}.
    \item Overtaking within the link is permitted, except in the queuing area, where lane changing is not allowed.
    \item The prior running time is assumed to follow a specific distribution during the study period.
\end{enumerate}

\section{Problem statement}\label{sec:problem_statement}
\begin{figure}[H]
    \centering
    \subfigure[A typical scenario for collecting LPR data at two adjacent intersections, illustrating overtaking behaviors in multiple lanes]{
        \includegraphics[width=0.8\textwidth]{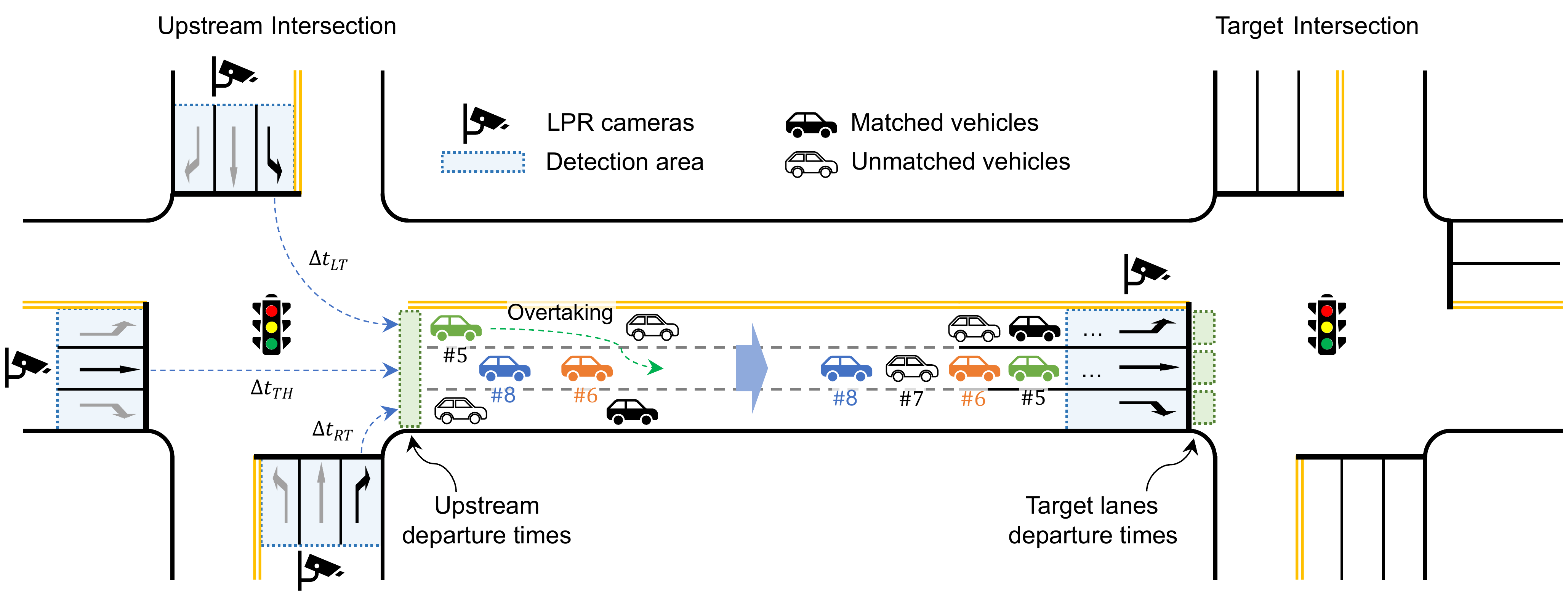}
        \label{fig:problem_1}
    }
    \hfill
    \subfigure[Cumulative arrival and departure curves of the through lane and queue profile estimation errors due to the FIFO assumption]{
        \includegraphics[width=0.8\textwidth]{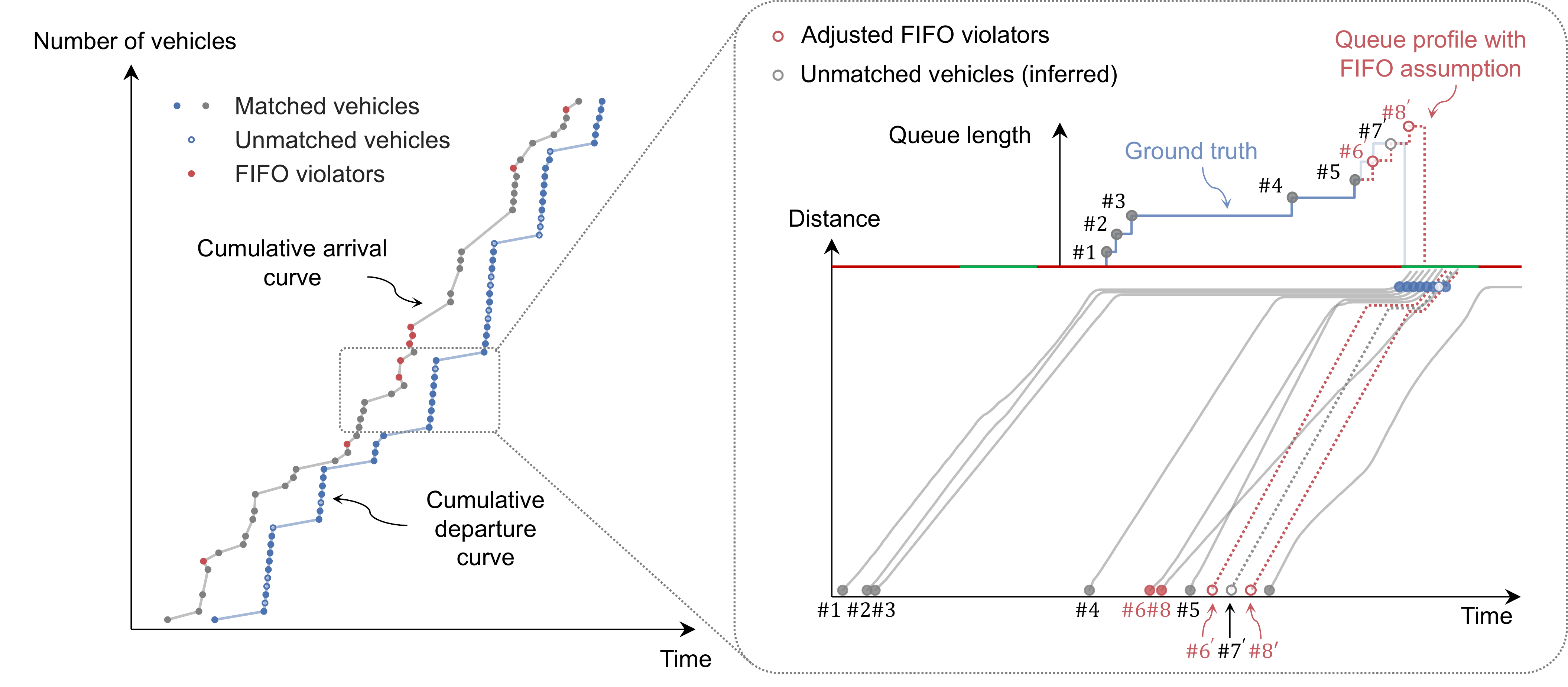}
        \label{fig:problem_2}
    }
    \caption{Problem statement.}
\end{figure}

As illustrated in Figure~\ref{fig:problem_1}, a typical scenario for collecting LPR data at two adjacent intersections is depicted. LPR cameras autonomously record vehicles passing the detection area, which spans 0-20 meters from the stop line and typically covers 1-3 vehicles. By combining signal timing information and travel time within the intersection, vehicles' detected times can be converted to departure times. The complexity of overtaking behaviors can lead to inconsistencies between the sequence of upstream departure times and the sequence of departure times for a specific lane. Figure~\ref{fig:problem_1} illustrates a possible vehicle travel process, where matched vehicles are represented by solid-filled icons and unmatched vehicles are represented by unfilled icons. For example, among vehicles departing from the through lane, vehicle 5 overtakes vehicles 6 and 8, whereas vehicle 7 fails to match with upstream departing vehicles due to errors in license plate number recognition. In this study, overtaking refers specifically to vehicles traveling to the same lane, which necessitates lane changing. This definition excludes overtaking due to speed differences between vehicles in different lanes, which is outside the scope of this research and unavailable from the LPR systems.

The detection results can be further described using cumulative arrival and departure curves, as shown in Figure~\ref{fig:problem_2}. Assuming no miss detection, the LPR camera of a specific lane (e.g., the through lane) can detect complete departures, as shown by the blue cumulative departure curve. In contrast, only partial arrivals can be observed, depending on the number of vehicles matched from the upstream intersection, as indicated by the gray cumulative arrival curve. The solid circles represent matched vehicles, while the hollow circles represent unmatched vehicles. The red solid circles denote FIFO violators resulting from overtaking, detailed in the partial vehicle trajectories and lane-based queue lengths in Figure~\ref{fig:problem_2}. This limitation makes it impossible to directly apply the input-output model to describe the traffic state of the link. Previous studies \citep{zhan_lane-based_2015, an_lane-based_2021} have addressed this issue by removing all FIFO violators. This would require adjusting their departure times to reasonable values based on certain rules (e.g., uniform arrival), and inferring the departure times of all unmatched vehicles. However, this adjustment process can introduce errors, specifically from incorrect inferred upstream departure times and neglecting interactions between vehicle trajectories.

\begin{figure}[H]
    \centering
    \includegraphics[width=0.75\linewidth]{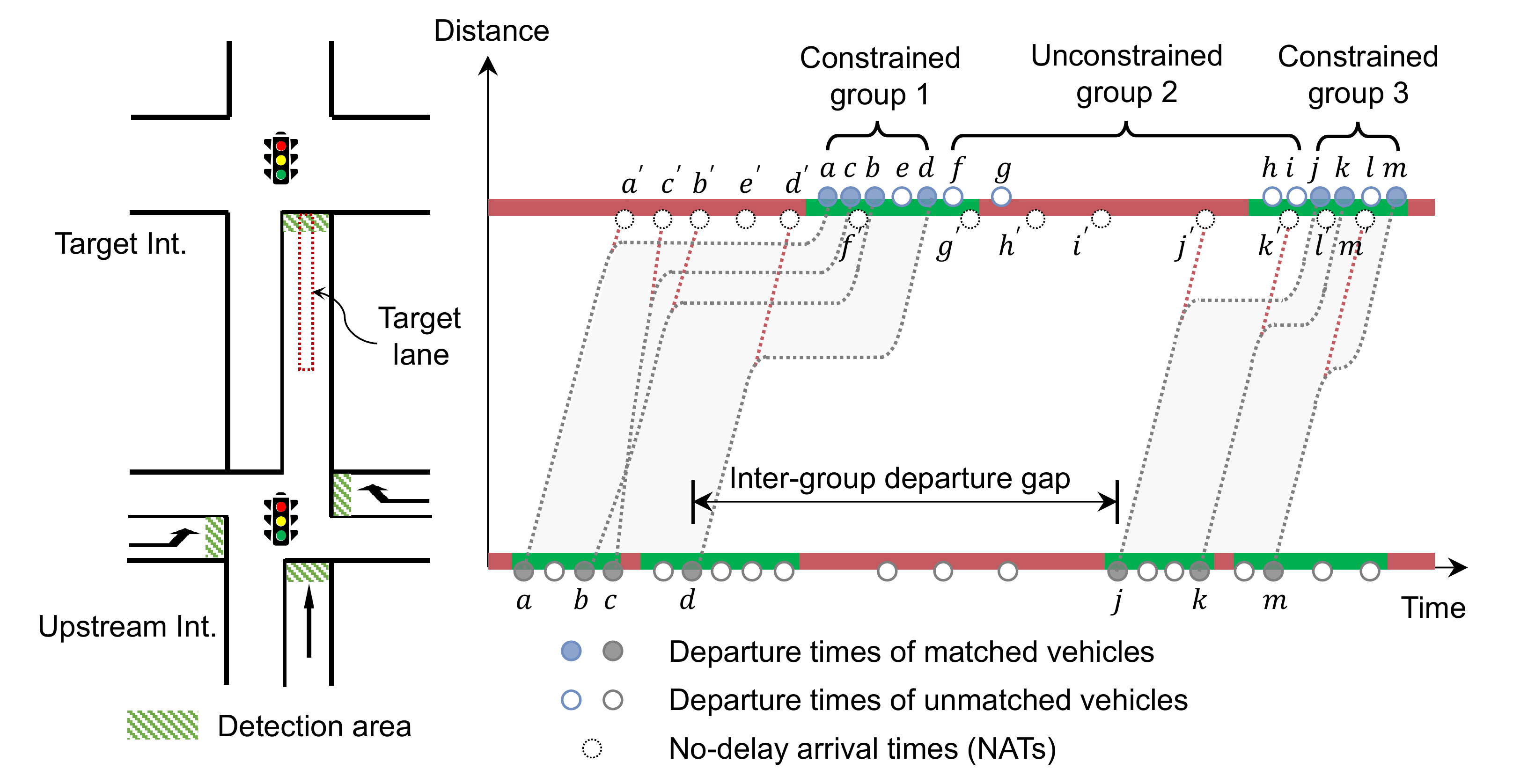}
    \caption{Illustration of the license plate matching result and no-delay arrival times (NATs).}
    \label{fig:nat}
\end{figure}

Instead of reconstructing the cumulative arrival and departure curves, which may incur issues such as FIFO violations, we focus on the vehicle's \textit{no-delay arrival time} (NAT), i.e., the arrival time without experiencing any delay. In Figure~\ref{fig:nat}, the NATs are denoted in the dashed hollow circle.
Given the specific NATs of all vehicles, the queue length can be easily derived by accumulating vehicles that queued. Thereby, the problem of queue length estimation is transformed into the estimation of NATs of vehicles. A straightforward way for NAT estimation is to assume a fixed running time for all vehicles, which ignores the vehicle speed variance and overtaking behaviors:

\begin{equation}
t^a_k = t^u_k + \tau_0, \quad \tau_0 \text{ is a constant}
\end{equation}

\noindent where $t^a_k$ is the NAT of vehicle $k$, $t^u_k$ is the upstream departure time of vehicle $k$, and $\tau_0$ is a fixed running time. Obviously, this approach introduces significant errors.

\citet{wu_stochastic_2024} considered the distribution of running time to better characterize the uncertain vehicle behaviors on the link and traversed possible overtaking cases. It only used the information provided by upstream LPR data, which can be summarized as

\begin{equation}
t^a_k = t^u_k + \tau_k, \quad \tau_k \sim \mathcal{D}(\cdot | \theta_{up})
\label{eq:wu}
\end{equation}

\noindent where $\tau_k$ is the running time of vehicle $k$, $\mathcal{D}(\cdot | \theta_{up})$ represents the distribution of $\tau$, and $\theta_{up}$ is the upstream information used to generate possible overtaking cases. 

However, Equation~\ref{eq:wu} ignores the actual overtaking information provided by the target intersection's LPR data. In Figure~\ref{fig:nat}, vehicles that departed during the first cycle from the target lane followed the order of $a$, $c$, $b$, $e$, $d$, among which vehicle $c$ arrived before $b$ due to overtaking. As lane-changing was not permitted in the queuing area, the sequence of NATs mirrors that of departure times. Subsequently, we can utilize the prior running time distribution and the license plate matching result (using information from both the upstream and the target intersection) to establish the NAT conditions, thereby indirectly considering the impact of overtaking. This process can be expressed as

\begin{equation}
t^a_k = t^u_k + \tau_k, \quad \tau_k \sim \mathcal{D}(\cdot | \theta_{up}, \theta_{target})
\end{equation}

\noindent where $\theta_{target}$ is the information from the target intersection.

Since we only utilize LPR data from a single lane of the target intersection in this scenario, we refer to this queue length estimation approach as the \textbf{single-lane estimation}. Notably, the single-lane estimation neglects information from upstream unmatched vehicles, which may depart from any lane of the target intersection. To leverage this information, we can build upon the single-lane estimation approach by identifying an optimal matching between unmatched vehicles departing from the upstream intersection and target lanes, thereby achieving \textbf{multi-lane estimation} and reducing the uncertainty in queue length estimation.

\section{Methodology}\label{sec:methodology}
\subsection{Framework}\label{sec:framework}
\begin{figure}[H]
    \centering
    \includegraphics[width=0.8\linewidth]{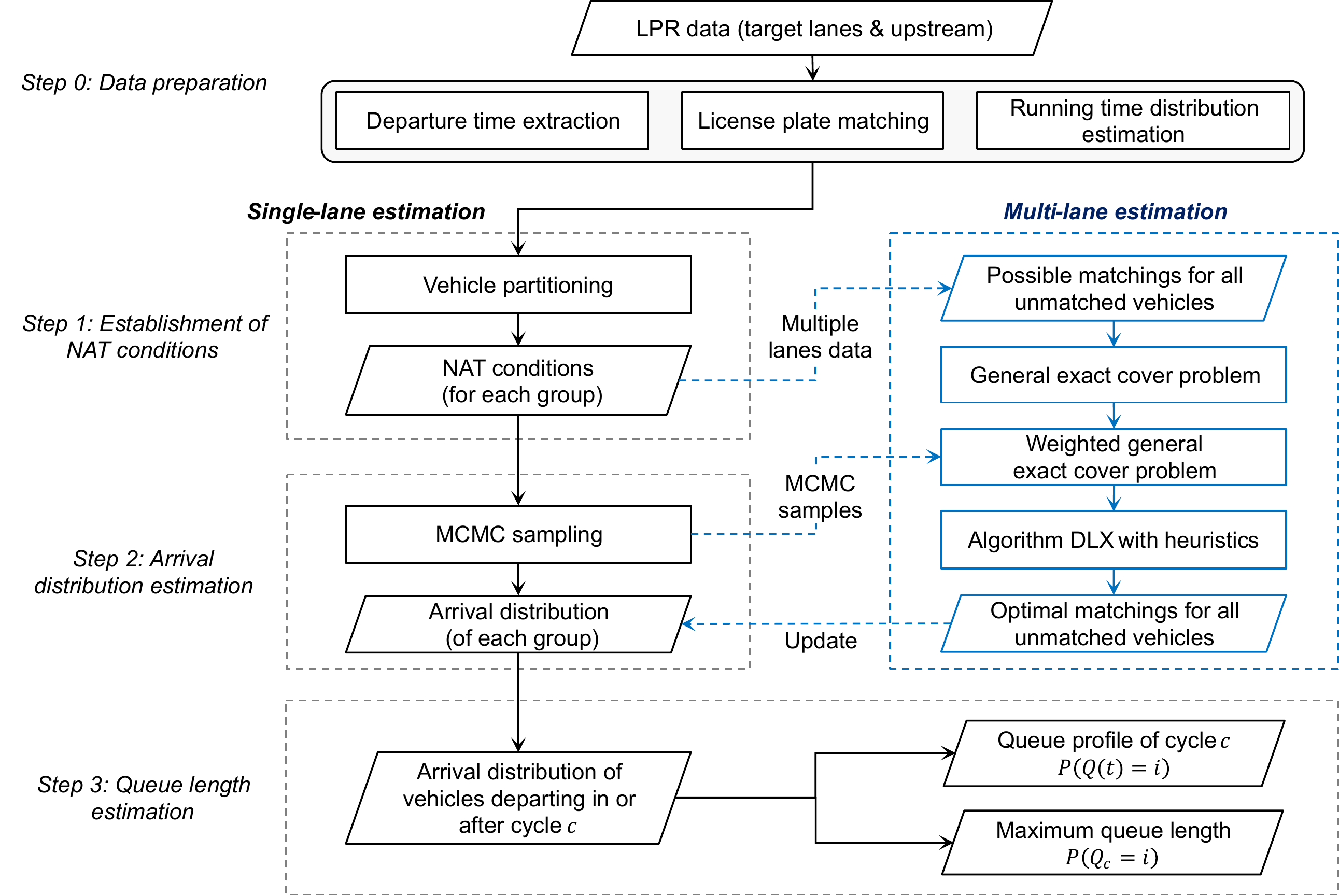}
    \caption{The general framework of the methodology.}
    \label{fig:framework}
\end{figure}

The general framework of the methodology is shown in Figure~\ref{fig:framework} and explained below.

In the \textbf{single-lane estimation}, we first process the LPR data to obtain data used in the next steps, including departure times, license plate matching results, and the running time distribution. Second, we can list the conditions all vehicles passing through the target lane must satisfy in NAT. Due to the large number of NAT conditions during the study period, we propose a DP-based vehicle partitioning algorithm to divide all vehicles passing through the target lane into constrained and unconstrained groups. The NAT conditions are then divided accordingly. Third, uniform samples of the region represented by the conditions can be obtained for each set of NAT conditions using an MCMC sampling algorithm. The conditional probability can be used to calculate the arrival distribution of all vehicles by integral calculation. Finally, the queue profile and maximum queue length for each cycle can be obtained by analyzing the arrival distribution of vehicles departing in or after each cycle in sequence.

In the \textbf{multi-lane estimation}, we extend the approach by using NAT conditions of multiple lanes and then update the arrival distribution. To achieve this, we first obtain possible matchings for each unmatched vehicle with upstream unmatched vehicles, and the problem of finding all possible matchings can be formulated as a general exact cover problem. Utilizing the uniform samples from the MCMC sampling, the problem can be reformulated to a weighted version and solved by the Algorithm DLX (an implementation of Knuth's Algorithm X \citep{knuth_dancing_2000} via dancing links) with heuristics. Finally, most previously unmatched vehicles will have been matched with upstream vehicles, enabling a more accurate computation of the arrival distribution and an improved estimation of queue profile and maximum queue length.

\subsection{Data preparation}

\subsubsection{Departure time extraction}
The LPR system provides only the detected times of vehicles. For vehicles detected during the red phase, their detected times do not represent their actual departure times. The detected time can be adjusted to determine the departure time using the existing interpolation method \citep{mo_speed_2017}. Furthermore, when considering the upstream intersection, an additional time component must be accounted for, which is the time it takes for a vehicle to leave the stop line and enter the link. This time varies depending on the movement type, with denotations of $\Delta t_{LT}$, $\Delta t_{TH}$, and $\Delta t_{RT}$ for left-turn, through, and right-turn movements, respectively, as illustrated in Figure~\ref{fig:problem_1}.

\subsubsection{License plate matching}
The license plate matching process consists of three steps. First, an outer join is executed on the LPR data from both the target lane and the upstream intersection, using the license plate number as a common key. Next, the difference between the downstream detected time and the upstream departure time, namely the travel time, is computed. Finally, records with travel times that are either less than 0 or exceed a certain threshold (e.g., 5 minutes) are filtered out.

\subsubsection{Running time distribution estimation}
To model the running time distribution, we employ the approach proposed by \citet{li_traffic_2023}, which involves fitting the travel time data using a mixed model with a log-normal distribution. After removing the noise component, the component with the smallest mean value is identified as the representative distribution of running time. Given that running times are inherently bounded, we truncate the log-normal distribution using the maximum and minimum values of travel time within this component:

\begin{equation}
    g(\tau;\mu,\sigma,a,b)=
    \begin{cases}
        \frac{\frac{1}{\tau\sigma}\phi(\frac{\ln\tau-\mu}{\sigma})}{\Phi(\frac{\ln b-\mu}{\sigma})-\Phi(\frac{\ln a-\mu}{\sigma})}, & a \leq \tau \leq b \\
        0, & \text{otherwise}
    \end{cases}
\end{equation}

\noindent Here, $g(\tau;\mu,\sigma,a,b)$ represents the probability density function (PDF) of the running time. The parameters $\mu$ and $\sigma$ denote the location and scale parameters of the log-normal distribution, respectively. The bounds $a$ and $b$ define the truncation interval, which corresponds to the minimum and maximum possible running times. Finally, $\phi(x)$ and $\Phi(x)$ represent the standard normal distribution's PDF and CDF (cumulative distribution function), respectively.

\subsection{Single-lane estimation}\label{sec:single_lane}
\subsubsection{Establishment of NAT conditions}\label{sec:NAT-conditions}
Given the information derived from the LPR data, we can establish the following conditions for all vehicles departing from the target lane:

\textit{1. Departure sequence conditions}

These conditions indicate that the sequence of the NATs must correspond to the sequence of departure times. If vehicle $k+1$ departs after vehicle $k$, then the NAT of vehicle $k+1$ should also be greater than that of vehicle $k$. Meanwhile, the headway condition should be satisfied: 

\begin{equation}
t_{k+1}^a - t_{k}^a \geq \min(h, t_{k+1}^d - t_{k}^d),\quad\forall k, k+1 \in \mathcal{K} \cup \mathcal{K}_u \label{eq:departure-sequence}
\end{equation}

\noindent where $\mathcal{K}$ is the set of vehicles in constrained groups; $\mathcal{K}_u$ is the set of vehicles in unconstrained groups; $t_{k}^a$ is the NAT of vehicle $k$; $t_k^d$ is the detected time of vehicle $k$; $h$ is the predefined saturation headway; $t_{k+1}^d - t_k^d$ is the actual arrival headway.

\textit{2. Running time conditions}

The running time of each vehicle follows a certain distribution and has a minimum and maximum value. Therefore, for all matched vehicles, we have

\begin{equation}
t_k^u + \tau_{k,min} \leq t_k^a \leq t_k^u + \tau_{k,max},\quad\forall k \in \mathcal{M} \label{eq:travel-time}
\end{equation}

\noindent where $\mathcal{M}$ is the set of matched vehicles; $t_k^u$ is the departure time from the upstream of vehicle $k$; $\tau_{k,min}$ and $\tau_{k,max}$ are the minimum and maximum running time for vehicles $k$, respectively.

\textit{3. Detected time conditions}

For all vehicles, the NATs should not exceed the actual detected times:

\begin{equation}
t_k^a \leq t_k^d,\quad\forall k \in \mathcal{K} \cup \mathcal{K}_u \label{eq:arrival-time}
\end{equation}

As mentioned, the arrival distribution estimation is conducted through an MCMC sampling method. The computational efficiency is greatly affected by the dimensionality of the sampling space, which is determined by the number of vehicles. Therefore, we need to partition the vehicles into several groups before sampling. The partitioning principle ensures that vehicles in adjacent constrained groups do not influence each other while maximizing the number of groups. This implies fewer vehicles in each group, thereby increasing the sampling efficiency. A DP-based algorithm, as shown in Algorithm~\ref{algo:partition}, was proposed to achieve this.

\begin{algorithm}[]
\caption{Matched vehicle partitioning}\label{algo:partition}
\KwIn{A list $lst$ of upstream departure times for matched vehicles, and a minimum inter-group departure gap $min\_gap$}
\KwOut{A list of partitions of matched vehicles}
Initialize a list $dp$ of size $n$ with all elements set to 1\;
Initialize a list $break\_points$ of size $n+1$ with all elements set to 0\;
$break\_points[dp[0]] \gets 1$\;
\For{$i \gets 1$ to $n-1$}{
    \For{$j \gets 0$ to $i-1$}{
        \If{$\min(lst[j+1:i+1]) - \max(lst[:j+1]) > min\_gap$}{
            $dp[i] \leftarrow \max(dp[j] + 1, dp[i])$;
        }
    }
    $break\_points[dp[i]] \leftarrow i + 1$;
}
Initialize an empty list $partitions$\;
\For{$i \gets 0$ to $dp[-1]-1$}{
    append $lst[break\_points[i]:break\_points[i+1]]$ to $partitions$;
}
\Return{$partitions$}
\end{algorithm}

Given that matched vehicles must be in constrained groups, we first select all matched vehicles, sorted by their arrival order, and use their upstream departure times as the first input, denoted $lst$. Next, we set the minimum inter-group departure gap $min\_gap$ to separate adjacent constrained groups, forming the second input. Due to the occurrence of overtaking, $lst$ may not necessarily increase monotonically. Consequently, the condition for non-influence becomes the form of $\min(lst[j+1:i+1]) - \max(lst[:j+1]) > min\_gap$, which implies that a sufficient amount of time has passed since all vehicles in the previous constrained group have departed from the upstream before vehicles in the current group begin to depart from the upstream. The list $dp$ is employed to store the maximum number of partitions that can be made up to a specific vehicle in the list, while the list $break\_points$ retains the indices in $lst$ where a new partition starts. The state transition equation $dp[i] = \max(dp[j] + 1, dp[i])$ is used to initiate a new partition if the non-influence condition is satisfied. Finally, we can derive the final partitioning results for matched vehicles according to the records in $break\_points$.

The classification of vehicles into constrained and unconstrained groups is determined by the partitioning result of matched vehicles. Specifically, each partition of matched vehicles, together with the unmatched vehicles departing from the target lane between them, forms a constrained group. In contrast, the unmatched vehicles partitioned by constrained groups are considered unconstrained groups. As shown in Figure~\ref{fig:nat}, for instance, vehicles $a$, $c$, $b$, $e$, $d$ along with vehicles $j$, $k$, $l$, $m$ form two constrained groups, whereas vehicle $f$, $g$, $h$, $i$ form an unconstrained group. This approach enables an initial estimation of the arrival distribution for vehicles in constrained groups, which serves as a foundation for estimating that for vehicles in unconstrained groups, thereby maximizing the utilization of matched LPR data.

After partitioning vehicles departing from the target lane, the preceding conditions from Equation~\ref{eq:departure-sequence} to~\ref{eq:arrival-time} can be assembled in matrix form. For each constrained group, these conditions can be expressed as

\begin{equation}
\begin{array}{r}
    \coolleftbrace{\makecell{\textit{1. Departure sequence} \\ \textit{conditions}}}{1 \\ 0 \\ \vdots \\ 0} \\
    \coolleftbrace{\makecell{\textit{2. Running time} \\ \textit{conditions}}}{1 \\ -1 \\ \vdots \\ 0 \\ 0} \\
    \coolleftbrace{\makecell{\textit{3. Detected time} \\ \textit{conditions}}}{1 \\ 0 \\ 0 \\ \vdots \\ 0}
\end{array}
\underbrace{
\begin{bmatrix}
    1 & -1 & 0 & \cdots & 0 \\
    0 & 1 & -1 & \cdots & 0 \\
    \vdots & \vdots & \ddots & \ddots & \vdots \\
    0 & 0 & \cdots & 1 & -1 \\
    1 & 0 & 0 & \cdots & 0 \\
    -1 & 0 & 0 & \cdots & 0 \\
    \vdots & \vdots & \vdots & \vdots & \vdots \\
    0 & 0 & 0 & \cdots & 1 \\
    0 & 0 & 0 & \cdots & -1 \\
    1 & 0 & 0 & \cdots & 0 \\
    0 & 1 & 0 & \cdots & 0 \\
    0 & 0 & 1 & \cdots & 0 \\
    \vdots & \vdots & \vdots & \ddots & \vdots \\
    0 & 0 & 0 & \cdots & 1
\end{bmatrix}
}_{\text{Matrix }\mathbf{A}}
\cdot
\underbrace{
\begin{bmatrix}
    t_1^a\\
    t_2^a\\
    t_3^a\\
    \vdots\\
    t_n^a
\end{bmatrix}
}_{\text{Vector }\mathbf{t}}
\leq
\underbrace{
\begin{bmatrix}
    -\min(h, t_{2}^d - t_{1}^d)\\
    -\min(h, t_{3}^d - t_{2}^d)\\
    \vdots\\
    -\min(h, t_{n}^d - t_{n-1}^d)\\
    t_1^u + \tau_{1,max}\\
    -t_1^u - \tau_{1,min}\\
    \vdots\\
    t_n^u + \tau_{n,max}\\
    -t_n^u - \tau_{n,min}\\
    t_1^d\\
    t_2^d\\
    t_3^d\\
    \vdots\\
    t_n^d
\end{bmatrix}
}_{\text{Vector }\mathbf{b}}
\label{eq:matrix-form}
\end{equation}

\noindent In this representation, $\mathbf{A}$ is an $m\times n$ matrix, where $m$ is the number of conditions and $n=|K|$ is the number of vehicles in the constrained group $K \in \mathcal{K}$. $\mathbf{t}$ represents an $n$-dimensional NAT vector, and $\mathbf{b}$ is an $m$-dimensional constant vector. Furthermore, an existing redundant constraints identification method \citep{telgen_identifying_1983} can be utilized to simplify the number of conditions.

As for unconstrained groups, it is impossible to determine the range of their NATs individually due to the lack of running time conditions. Therefore, it is necessary to consider the conditions of vehicles immediately previous and following the unconstrained group, as in the case of vehicle $d$ and vehicle $j$ depicted in Figure~\ref{fig:nat}. The conditions can be formulated as a similar matrix form like Equation~\ref{eq:matrix-form}. In this case, $\mathbf{A}$ is a $m\times (n+2)$ matrix, $\mathbf{t}$ represents an $(n+2)$-dimensional NAT vector, and $\mathbf{b}$ is an $m$-dimensional constant vector.

\subsubsection{Arrival distribution estimation}

\paragraph{Constrained group}
After establishing NAT conditions for each vehicle group, we first compute the arrival distributions of vehicles in each constrained group. Assuming we know the prior probability distributions of running times for matched vehicles, we can then derive the prior probability distribution of their NATs, given that their upstream departure times have been captured by the LPR camera. Due to the lack of upstream information for unmatched vehicles, we assume that their prior NATs follow a uniform distribution. If we consider the NATs of all vehicles as independent variables, their joint PDF can be calculated through multiplication. Under the conditions obtained in the previous step, the probability of each vehicle's NAT falling in $[t,t+1]$ can be calculated through the conditional probability:

\begin{equation}
   \begin{aligned}
   P(t\leq t_k^a \leq t+1 | \mathbf{A} \mathbf{t} \leq \mathbf{b})&=
    \frac{P((\mathbf{A}^T,\mathbf{A}_{k,t}^T)^T\mathbf{t}\leq(\mathbf{b}^T,\mathbf{b}_{k,t}^T)^T)}{P(\mathbf{A}\mathbf{t}\leq \mathbf{b})} \\
    &=\frac{\int_{\Omega_{k,t}}f(\mathbf{t})d\mathbf{t}}{\int_{\Omega}f(\mathbf{t})d\mathbf{t}},\quad k \in K \label{eq:conditional}
  \end{aligned}
\end{equation}

\noindent where $\mathbf{A}_{k,t}\mathbf{t}\leq \mathbf{b}_{k,t}$ indicates that vehicle $k$'s NAT falls in $[t,t+1]$, which is the matrix representation of $t\leq t_k^a \leq t+1$. $\Omega=\{\mathbf{t} \in \mathbb{R}^n|\mathbf{A} \mathbf{t} \leq \mathbf{b}\}$ and $\Omega_{k,t}=\{\mathbf{t} \in \mathbb{R}^n|(\mathbf{A}^T,\mathbf{A}_{k,t}^T)^T\mathbf{t}\leq(\mathbf{b}^T,\mathbf{b}_{k,t}^T)^T\}$ are polytopes composed of linear inequality constraints. $f(\mathbf{t})=\prod_{k\in K} f_k(t_k^a)$ is the joint PDF of $\mathbf{t}$, and for matched and unmatched vehicles, we have

\begin{equation}
    f_k(t_k^a)=
    \begin{cases}
        g_k(t_k^a-t_k^u), & k\in K \cap \mathcal{M} \\
        constant, & k \in K - \mathcal{M}
    \end{cases}
\end{equation}

\noindent where $g_k(\tau_k)$ represents the PDF of vehicle $k$'s running time, denoted as $\tau_k = t_k^a - t_k^u$. For unmatched vehicles, their prior PDFs for NATs are the same for each second. However, this specific value does not impact subsequent computations and can thus be ignored when calculating joint PDF $f(\mathbf{t})$.

Since the dimension of the vector $\mathbf{t}$, $n=|K|$, is usually high, it is challenging to compute using deterministic integration algorithms. Hence, we employ the naive Monte Carlo integration to calculate the above probabilities. For integrals in the form $\int_{\Omega}f(\mathbf{t})d\mathbf{t}$, we first calculate the volume of polytope $\Omega$:

\begin{equation}
    V=\int_{\Omega}d\mathbf{t}
\end{equation}

Next, we uniformly sample $\mathbf{t}_1,\mathbf{t}_2,...,\mathbf{t}_N \in \Omega$ within the polytope $\Omega$. We chose an MCMC sampling algorithm called Vaidya walk \citep{chen_fast_2018} to achieve that. Vaidya walk is a random walk derived from interior point methods and based on the volumetric-logarithmic barrier introduced by \citet{vaidya_new_1989}. For a polytope in $\mathbb{R}^n$ defined by $m$ constraints, the Vaidya walk mixes in $\mathcal{O}(m^{0.5}n^{1.5})$ steps, with an effective cost of $\mathcal{O}(m^{1.5}n^{3.5})$ per sample. Since $m\ge n$ for closed polytopes in $\mathbb{R}^n$, the Vaidya walk is more efficient than other random walks, such as the Dikin walk \citep{kannan_random_2012}, which has an effective cost of $\mathcal{O}(m^{2}n^{3})$ per sample. The samples obtained are then used to compute the integral value, which approximates the average joint PDF values of the samples multiplied by the volume of the polytope:

\begin{equation}
    \int_{\Omega} f(\mathbf{t}) d \mathbf{t} \approx V \frac{1}{N} \sum_{i=1}^N f(\mathbf{t}_i)\label{eq:mcmc_integral}
\end{equation}

Since vehicles must arrive and can only arrive once, any point uniformly sampled in $\Omega$ will always fall into one of the $\Omega_{k,t}$. Let's denote the points that fall into $\Omega_{k,t}$ as $\mathbf{t}_{k,t,1},\mathbf{t}_{k,t,2},...,\mathbf{t}_{k,t,N_{k,t}}\in \Omega_{k,t}$, with a count of $N_{k,t}$. Then, approximately, we have

\begin{equation}
    \frac{V_{k,t}}{V} \approx \frac{N_{k,t}}{N}
\end{equation}

\noindent Here, $V_{k,t}$ is the volume of $\Omega_{k,t}$. Therefore, the conditional probability in Equation~\ref{eq:conditional} can be expressed as the ratio of the sums of the joint PDF values of the sampling points, where the difficult-to-calculate volume can be canceled out in the ratio:

\begin{equation}
    \frac{\int_{\Omega_{k,t}}f(\mathbf{t})d\mathbf{t}}{\int_{\Omega}f(\mathbf{t})d\mathbf{t}}\approx
    \frac{V_{k,t}\frac{1}{N_{k,t}}\sum_{i=1}^{N_{k,t}}f(\mathbf{t}_{k,t,i})}
    {V\frac{1}{N}\sum_{i=1}^{N}f(\mathbf{t}_i)}\approx
    \frac{\sum_{i=1}^{N_{k,t}}f(\mathbf{t}_{k,t,i})}
    {\sum_{i=1}^{N}f(\mathbf{t}_i)}\label{eq:ratio}
\end{equation}

The procedure to calculate the arrival distribution is summarized in Algorithm~\ref{algo:mhar}.

\begin{algorithm}[]
\caption{Arrival distribution estimation}\label{algo:mhar}
\KwIn{Vehicle group $K$ along with the corresponding NAT conditions, denoted as $\mathbf{A}\mathbf{t}\leq \mathbf{b}$}
\KwOut{Arrival distribution of each vehicle}
Uniformly sample $\mathbf{t}_1, \mathbf{t}_2, ..., \mathbf{t}_N \in \Omega=\{\mathbf{t} \in \mathbb{R}^n|\mathbf{A} \mathbf{t} \leq \mathbf{b}\}$ and denote $\mathbf{T} = (\mathbf{t}_1, \mathbf{t}_2, ..., \mathbf{t}_N)$\;
Calculate $\sum_{i=1}^N f(\mathbf{t}_i)$\;
\ForEach{vehicle $k$ in $K$}{
    Compute the minimum value $t_{k,min}^a$ and the maximum value $t_{k,max}^a$ of $t_k^a$ based on $\mathbf{A}\mathbf{t}\leq \mathbf{b}$, then round these values down and up respectively\;
    \For{$t \gets t_{k,min}^a$ to $t_{k,max}^a$}{
        Select $N_{k,t}$ points from $\mathbf{T}$ which satisfy $\mathbf{A}_{k,t}\mathbf{t}\leq\mathbf{b}_{k,t}$\;
        Compute $\sum_{i=1}^{N_{k,t}} f(\mathbf{t}_{k,t,i})$ to get the approximation of Equation~\ref{eq:conditional} using Equation~\ref{eq:ratio}\;
    }
}
\end{algorithm}

\begin{figure}[H]
    \centering
    \includegraphics[width=0.75\linewidth]{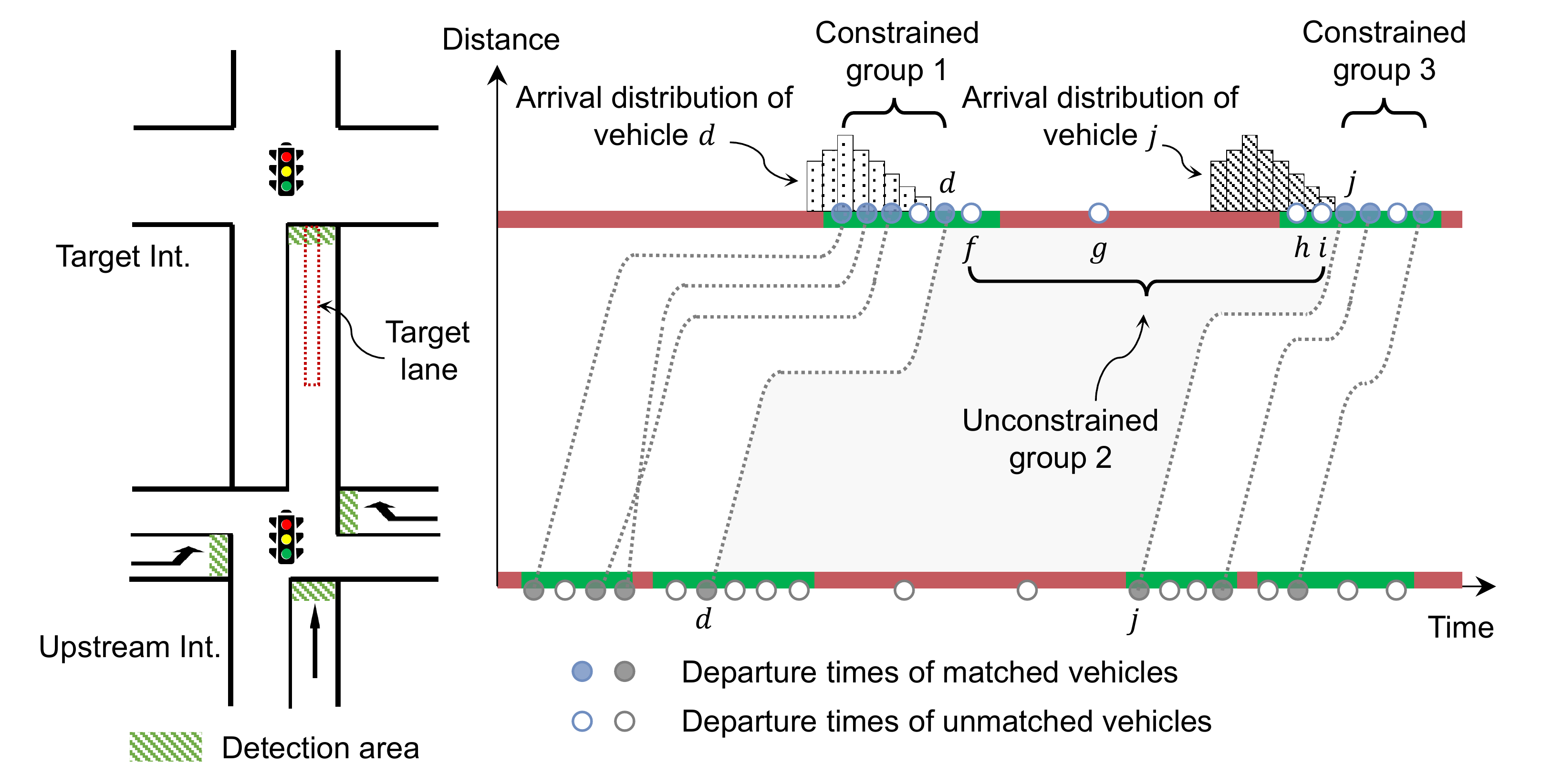}
    \caption{Illustration of the arrival distribution estimation for an unconstrained group.}
    \label{fig:unconstrained}
\end{figure}

\paragraph{Unconstrained group}
Following the calculation of the arrival distribution for adjacent constrained groups, each unconstrained group can be bounded. As outlined by Algorithm~\ref{algo:partition}, we ensure minimal interference between vehicles in distinct constrained groups, and then their NATs are deemed conditionally independent. In Figure~\ref{fig:unconstrained}, the vehicles $f$, $g$, $h$, $i$ constitute an unconstrained group, and the arrival distributions of both the preceding and following vehicles, i.e., vehicle $d$ and $j$, have already been obtained. The procedure to compute the arrival distributions of vehicles in an unconstrained group is similar to Algorithm~\ref{algo:mhar}, with the primary difference being the calculation of the joint PDF $f(\mathbf{t})$, as shown in Equation~\ref{eq:unconstrained-joint-pdf}. At this point, the arrival distributions of the previous and following vehicles are known, so the PDFs of their NATs can be treated as second-based piecewise functions, as demonstrated in the histogram in Figure~\ref{fig:unconstrained}:

\begin{equation}
f(\mathbf{t})=f_{prev.}(t_{prev.}^a)\cdot f_{foll.}(t_{foll.}^a)\label{eq:unconstrained-joint-pdf}
\end{equation}

\noindent where $f_{prev.}(t_{prev.}^a)$ represents the arrival distribution of the previous vehicle, and $f_{foll.}(t_{foll.}^a)$ signifies that of the following vehicle. Given that the remainder of the vehicles are all unmatched, their PDFs are disregarded.

\subsubsection{Queue length estimation}\label{sec:queue-length-estimation}
After obtaining the arrival distribution of all vehicles departing from the target lane, we can further calculate the queue profile and the maximum queue length per cycle.

\begin{figure}
    \centering
    \includegraphics[width=0.75\textwidth]{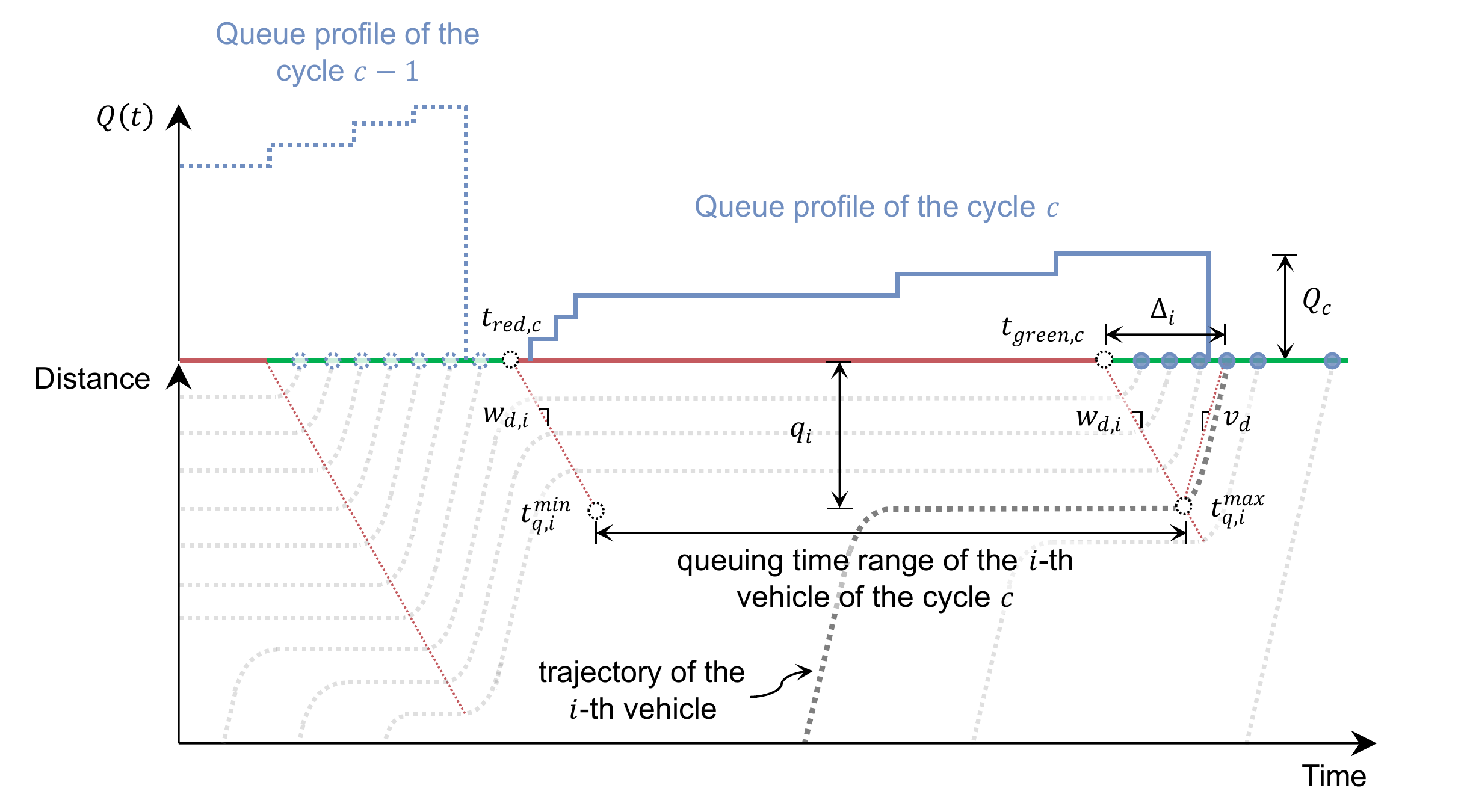}
    \caption{Illustration of the queuing time range of each vehicle per cycle.}
    \label{fig:queue_profile_estimation_1}
\end{figure}

To obtain the queue profile, we need to calculate the distribution of queue length $Q(t)$ at each time step $t$ based on the arrival distribution estimated in the previous section. First, we must determine the potential queuing time range for each vehicle within a cycle. Figure~\ref{fig:queue_profile_estimation_1} shows a typical cycle's vehicle trajectory, where cycle $c-1$ is over-saturated, resulting in a residual queue. For cycle $c$, the queuing time range for the $i$-th vehicle (whose trajectory is shown in dark gray in the figure) can be calculated using the dissipation wave speed $w_{d,i}$ before the $i$-th vehicle departs:

\begin{gather}
    t_{q,i}^{min} = t_{red,c} + \frac{q_i}{w_{d,i}} \\
    t_{q,i}^{max} = t_{green,c} + \frac{q_i}{w_{d,i}}
\end{gather}

\noindent Here, $t_{q,i}^{min}$ and $t_{q,i}^{max}$ represent the potential queuing time range in cycle $c$ for the $i$-th vehicle departing in or after cycle $c$. $t_{red,c}$ and $t_{green,c}$ are the start times of the red and green lights in cycle $c$, respectively. $q_i = \sum_{k=1}^{i} x_k$ is the queue length in meters if there are $i$ vehicles in the queue, where $x_k$ is the queuing space in meters of the $k$-th vehicle. To calculate $w_{d,i}$, we assume that vehicles travel at a fixed discharging speed $v_d$ before passing the stop line. According to the shockwave theory, we obtain:

\begin{gather}
    w_{d,i} = \frac{v_d \cdot q_i}{v_d \cdot \Delta_i - q_i} \\ 
    \Delta_i = \min\{h \cdot i, t_i^{dd} - t_{green, c}\}
\end{gather}

\noindent where $\Delta_i$ is the minimum time difference between the start time of the green light and when the $i$-th vehicle departs, $h$ is the saturation headway, and $t_i^{dd}$ is the actual departure time of the $i$-th vehicle. Notably, for vehicles appearing in multiple cycles, each cycle has a corresponding potential queuing time range.

\begin{figure}[H]
    \centering
    \subfigure[Queue profile estimation]{
        \includegraphics[width=0.48\textwidth]{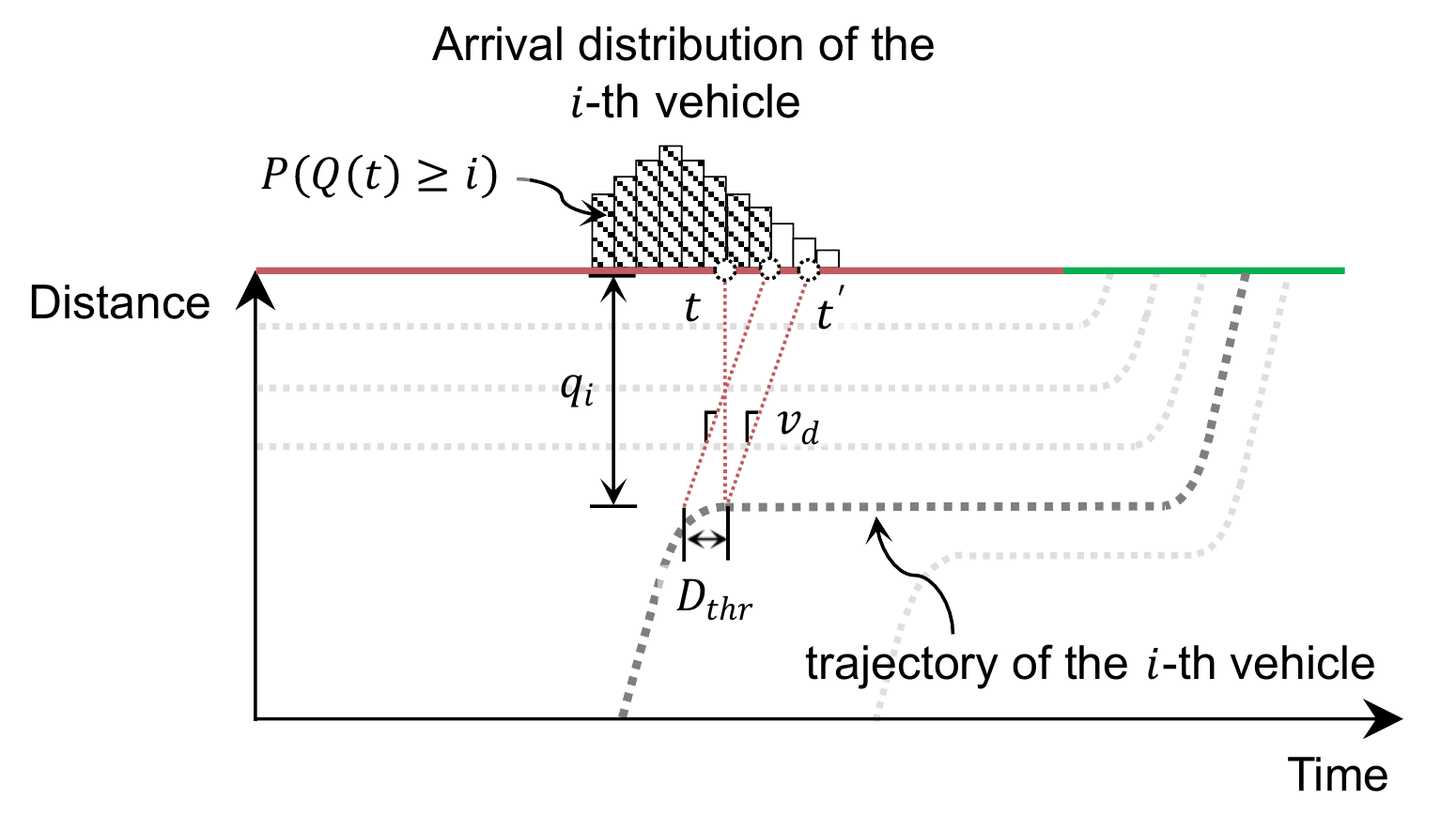}
        \label{fig:queue_profile_estimation_2}
    }
    \hfill
    \subfigure[Maximum queue length estimation]{
        \includegraphics[width=0.48\textwidth]{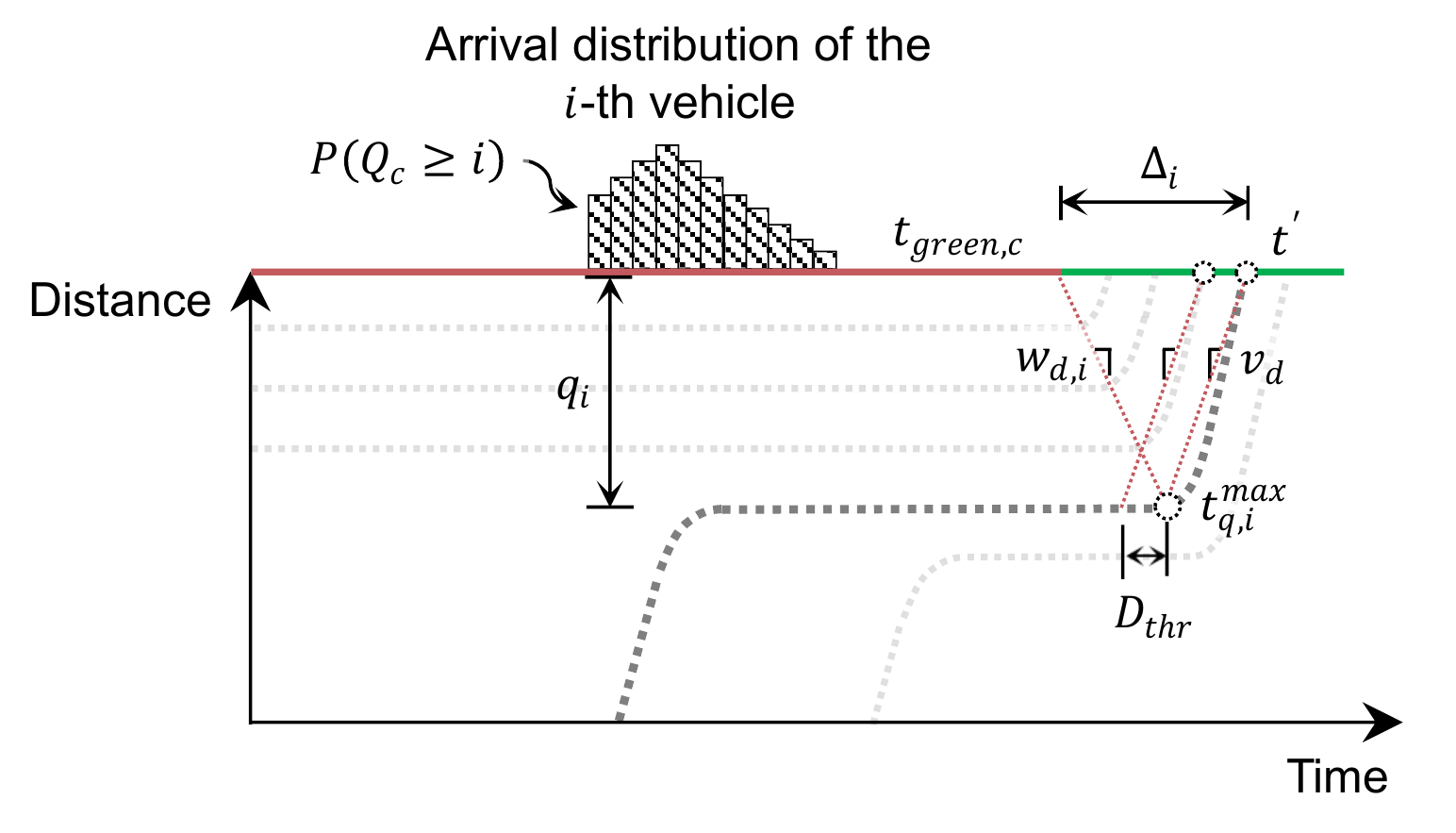}
        \label{fig:queue_profile_estimation_3}
    }
    \caption{Illustration of queue length estimation process.}
\end{figure}

For any time $t$, we can calculate the lower bound of the queue length at time $t$ based on the arrival distribution of all vehicles whose queuing time range covers that moment. As shown in Figure~\ref{fig:queue_profile_estimation_2}, the underlying concept is that if the $i$-th vehicle departing in or after cycle $c$ arrives at its queuing position before time step $t$, it can be inferred that the queue length is at least $i$. However, the previously estimated arrival distribution is based on arrivals at the stop line. Given the queue length in meters $q_i$ when there are $i$ vehicles in the queue, we can calculate the corresponding arrival time at the stop line:

\begin{equation}
    t' = t + \frac{q_i}{v_d}
\end{equation}

\noindent where $t$ is the current time step, and $t'$ is the time when the $i$-th vehicle is expected to arrive at the stop line.

To eliminate the impact of acceleration and deceleration delays, we set a delay threshold $D_{thr}$ to avoid treating minor fluctuations in arrival times as stopped delays. A vehicle is only considered to be queuing if its delay exceeds $D_{thr}$, i.e. if its NAT is before $t' - D_{thr}$. This enables us to calculate the probability that the queue length at time step $t$ equals or exceeds $i$:

\begin{equation}
    P(Q(t)\ge i) = P(t_i^a \le t' - D_{thr})
\end{equation}

Then the probability mass function (PMF) of $Q(t)$ can be obtained by subtraction:

\begin{equation}
    P(Q(t)=i - 1) = P(Q(t)\ge i - 1) - P(Q(t) \ge i)
    \label{eq:subtraction}
\end{equation}

However, for several time steps at the beginning of the green light, the vehicles before the $i$-th vehicle may no longer be queuing, and thus $P(Q(t)\ge i-1)$ cannot be obtained, making Equation~\ref{eq:subtraction} inapplicable. A typical example is as follows, where the $i$-th element represents the value of $P(Q(t)\ge i)$:

\begin{equation}
    [1, na, na, na, P(Q(t) \ge 4), P(Q(t) \ge 5), 0]
\end{equation}

\noindent In this case, the probabilities for queue lengths of 1, 2, and 3 are not available since the first, second, and third vehicles are not within their queuing time range. Therefore, we first fill the NA values by using the next valid estimation, i.e., the following relationship:

\begin{equation}
    P(Q(t) \ge 1) = P(Q(t) \ge 2) = P(Q(t) \ge 3) = P(Q(t) \ge 4)
\end{equation}

\noindent Then, Equation~\ref{eq:subtraction} can be simply calculated.

To estimate the maximum queue length for cycle $c$, we can bypass the estimation of the entire cycle's queue profile. As illustrated in Figure~\ref{fig:queue_profile_estimation_3}, for the $i$-th vehicle, the position with the highest queuing probability occurs at the maximum possible queuing time step $t_{q,i}^{max}$:

\begin{equation}
    P(Q_c \ge i) = P(Q(t_{q,i}^{max}) \ge i)
\end{equation}

\noindent Therefore, we only need to sequentially calculate the probability for all vehicles departing in or after cycle $c$ until $P(Q_c \ge i) = 0$. Then, similar to Equation~\ref{eq:subtraction}, we can compute the PMF of $Q_c$:

\begin{equation}
    P(Q_c=i-1) = P(Q_c \ge i-1) - P(Q_c \ge i)
\end{equation}

The queue profile and maximum queue length estimation processes are summarized in Algorithm~\ref{algo:queue_length}. Notably, if we only need to calculate the maximum queue length, the parameter $v_d$ is not required, as $t'$ can be directly determined by $\Delta_i$.

\begin{algorithm}[]
\caption{Estimation of the queue profile and the maximum queue length}\label{algo:queue_length}
\KwIn{The set $K_c$ of all vehicles departing in or after cycle $c$ and their corresponding information, the delay threshold $D_{thr}$, the saturation headway $h$, and the discharging speed $v_d$}
\KwOut{The queue profile $Q(t)=i$ and the maximum queue length $Q_c$}
Calculate the queue length in meters $q_i=\sum_{k=1}^{i} x_k$ if there are $i$ vehicles in the queue\;
\tcp{Queue profile estimation}
\ForEach{time step $t$ in cycle $c$}{
    $P(Q(t)\ge 0) = 1$\;
    \For{the $i$-th vehicle in $K_c$ satisfying $t_{q,i}^{min} \le t \le t_{q,i}^{max}$}{
        $t' \gets t + \frac{q_i}{v_d}$\;
        $P(Q(t) \ge i) = P(t_i^a \le t' - D_{thr})$\;
    }
    Fill the NA values in the sequence of $P(Q(t)\ge i)$ by using the next valid estimation\;
    Calculate $P(Q(t)=i-1) = P(Q(t)\ge i-1) - P(Q(t) \ge i)$ for each $i$ until $P(Q(t) \ge i)=0$\;
}
\tcp{Maximum queue length estimation}
\For{the $i$-th vehicle in $K_c$}{
    $t' \gets t_{green,c} + \Delta_i$\;
    $P(Q_c \ge i) = P(t_i^a \le t' - D_{thr})$\;
    \If{$P(Q_c \ge i) = 0$}{
        \textbf{break}\;
    }
}
Calculate $P(Q_c=i-1) = P(Q_c \ge i-1) - P(Q_c \ge i)$ for all iterated $i$\;
\end{algorithm}

\subsection{Multi-lane estimation}\label{sec:multi_lane}

\begin{figure}[H]
    \centering
    \includegraphics[width=0.75\linewidth]{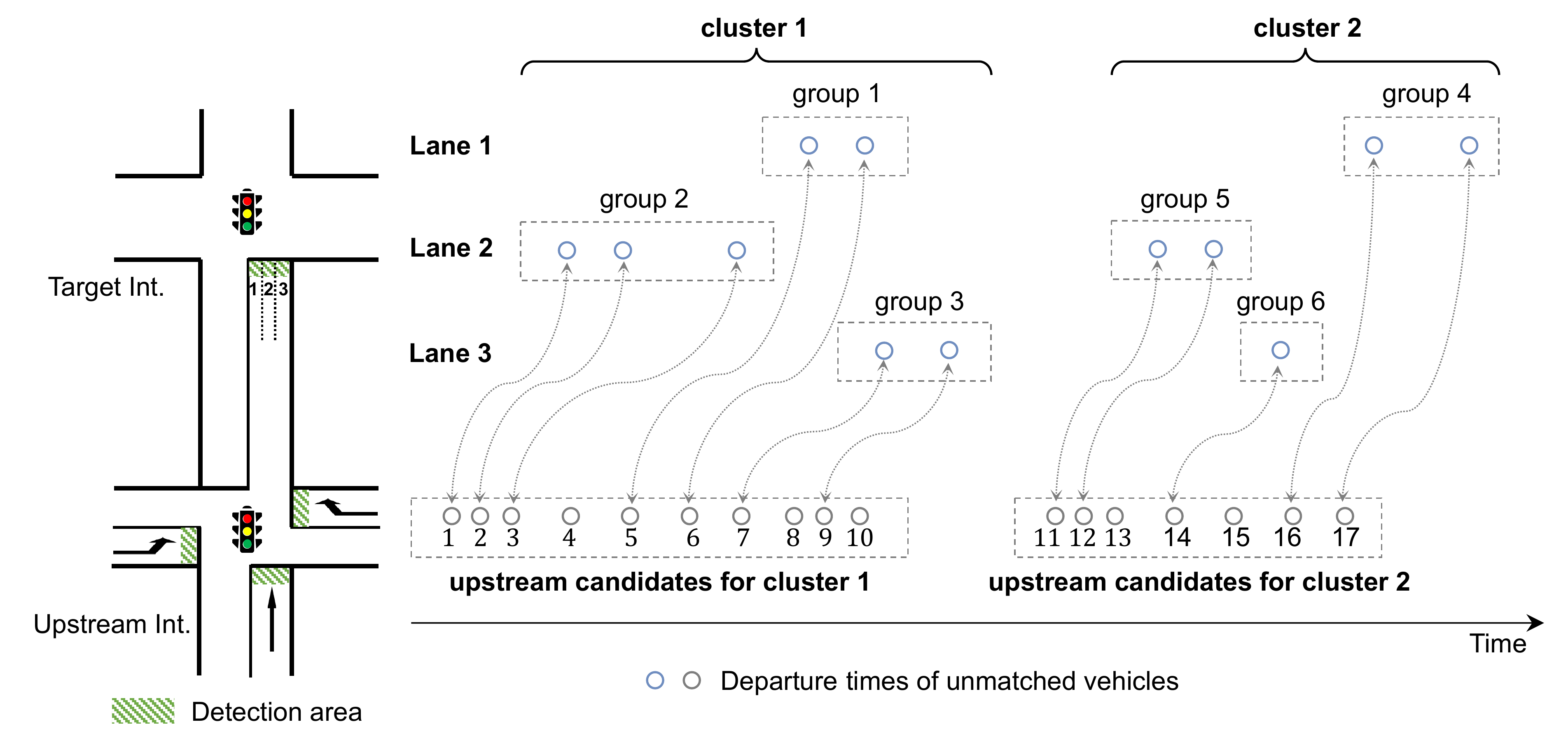}
    \caption{Illustration of possible matchings with upstream vehicles for each group.}
    \label{fig:multi_lane}
\end{figure}

For each lane in the approach, we have several groups (both constrained and unconstrained) that contain unmatched vehicles, as shown in Figure~\ref{fig:multi_lane}. For each unmatched vehicle, we can compute the minimum value $t_{k,min}^a$, and the maximum value $t_{k,max}^a$ of its NAT according to the NAT conditions established in the previous section. Then based on the minimum value $\tau_{k,min}$ and the maximum value $\tau_{k,max}$ of its running time, the departure time range of upstream vehicles that could match with it can be expressed as $[t_{k,min}^a - \tau_{k,max}, t_{k,max}^a - \tau_{k,min}]$. According to this range, we can obtain the upstream candidates for this vehicle. Subsequently, the upstream candidates for each group can be represented as the union of the upstream candidates of each vehicle within it. For all groups within the study period, if the union of upstream candidates of some groups has no intersection with the union of upstream candidates of other groups, then these groups can be classified into one cluster, such as groups 1, 2, 3, and groups 4, 5, 6 in Figure~\ref{fig:multi_lane}.

For individual vehicles, upstream candidates indicate possible matchings to upstream vehicles. For each group, the possible matchings are the combination of upstream candidates of each vehicle. For example, for group 2 in Figure~\ref{fig:multi_lane}, the upstream candidates for each vehicle are $[1]$, $[1,2,3]$, and $[2,3,4,5]$ respectively, then valid combinations include $[1,2,3]$, $[1,2,4]$, $[1,2,5]$, $[1,3,2]$, etc. However, combinations such as $[1,1,2]$ and $[1,3,3]$ are not permitted, as vehicles are duplicated. The figure also shows possible matchings for other groups. Our task is to find the optimal matching for each group within each cluster, that is, to determine the optimal global matching for each cluster.

\begin{remark}
When faced with an exponentially large number of combinations, sampling techniques can be employed to select a representative subset, thereby avoiding the computational burden of considering all possible combinations.
\end{remark}

\subsubsection{Problem transformation}

For each cluster, the problem of identifying all possible global matchings can be formulated as a general exact cover problem. Furthermore, considering the total weight associated with each global matching, this problem can be transformed into a weighted version. Before proceeding, let us briefly introduce the concept of the exact cover problem:

\begin{definition}[Exact cover problem]
Given a collection $S$ of subsets of a set $X$, the exact cover problem is to find a subcollection $S^*$ of $S$ that satisfies the following condition:

\begin{itemize}
    \item Each element in $X$ is contained in exactly one subset in $S^*$.
\end{itemize}
\end{definition}

\noindent The exact cover problem can be generalized slightly to involve not only exactly-once constraints but also at-most-once constraints:

\begin{definition}[General exact cover problem]
Given a collection $S$ of subsets of a set $X$, the general exact cover problem is to find a subcollection $S^*$ of $S$ that satisfies the following conditions:

\begin{itemize}
\item Each element in a subset $X^* \subseteq X$ is contained in at most one subset in $S^*$.
\item Each element in $X \setminus X^*$ is contained in exactly one subset in $S^*$.
\end{itemize}
\end{definition}

\begin{figure}[H]
    \centering
    \includegraphics[width=0.75\linewidth]{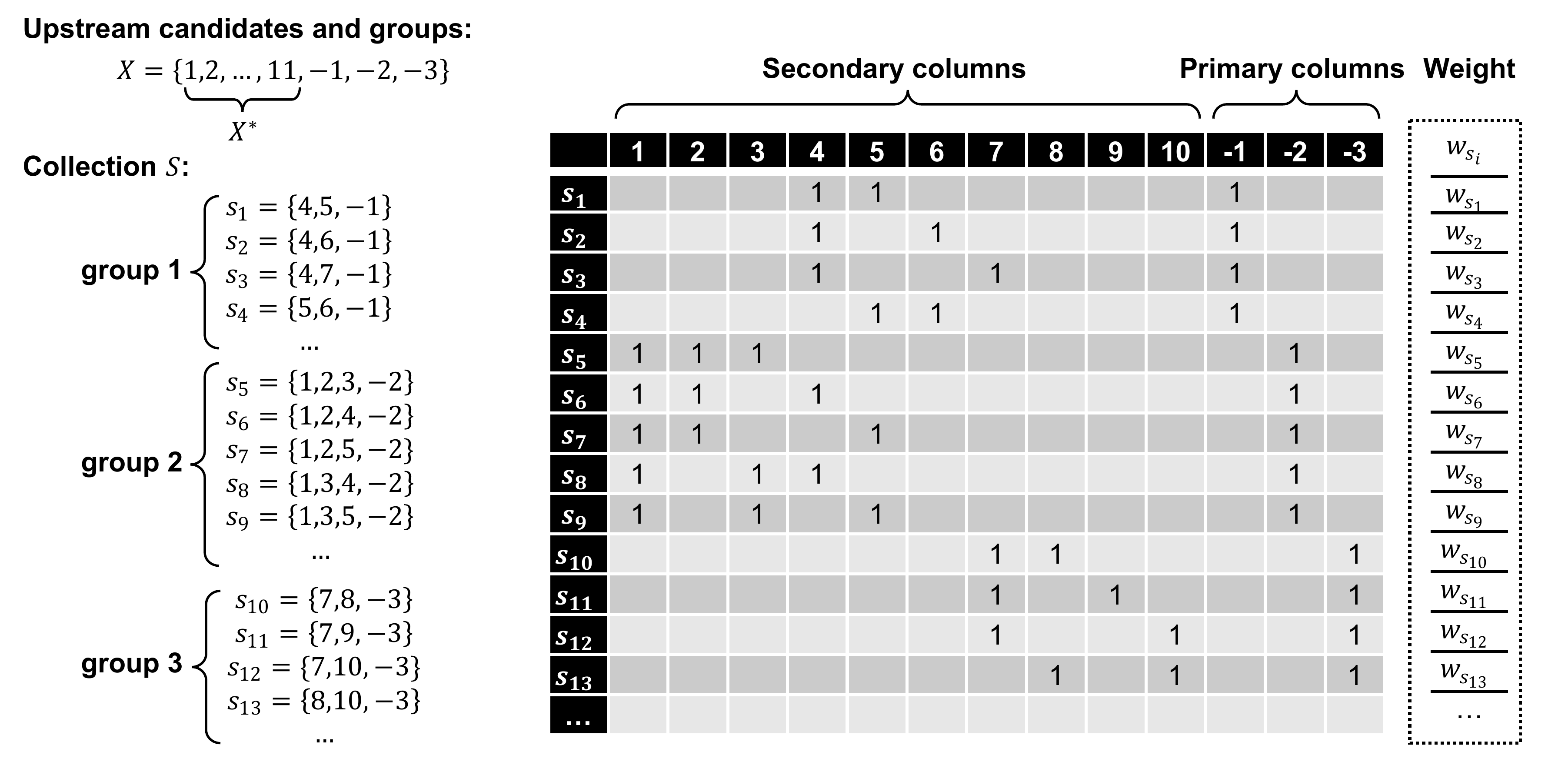}
    \caption{Illustration of the weighted general exact cover problem to find the optimal global matching.}
    \label{fig:exact_cover}
\end{figure}

Figure~\ref{fig:exact_cover} illustrates the general exact cover problem for finding a possible global matching for cluster 1. Set $X$ includes all the numbers of upstream candidates as well as the numbers of groups (represented by negative numbers). Among these, the numbers of upstream candidates make up the subset $X^*$, meaning that upstream unmatched vehicles are not necessarily matched with any group, but may instead leave the link midway. For example, in Figure~\ref{fig:multi_lane}, the number of upstream candidates for cluster 1 is 10, but the total number of vehicles in all groups is only 7. Subsets in collection $S$ define the possible matchings for different groups. Since sets are inherently unordered, a subset $s_i$ may correspond to multiple matchings for group $j$, denoted as $\Theta_j(s_i)$. For instance, the two unmatched vehicles in group 1 can be matched with upstream vehicles 5 and 6, or 6 and 5, both of which can be represented by the subset $s_1=\{5,6,-1\}$.

For clearer explanation, the subsets can be expressed in matrix form as shown on the right side of Figure~\ref{fig:exact_cover}. Each row represents a subset of set $X$, and the column corresponding to the position of 1 is the element of that subset. Columns corresponding to subset $X^*$ are called secondary columns, while all others are referred to as primary columns. Therefore, our task is to select several rows from the matrix such that primary columns contain exactly one 1, and secondary columns contain at most one 1. It is easy to see that $S^*=\{s_1,s_5,s_{10}\}$, $S^*=\{s_2,s_7,s_{11}\}$, and so on, are possible solutions.

Furthermore, to evaluate different solutions, we reformulated the problem to find an optimal global matching as a maximum likelihood estimation (see Appendix~\ref{sec:appendixA}). The global match's log-likelihood function $l(\theta)$ can be represented as the sum of the log-likelihood functions of different groups $l_j(\theta_j)$:

\begin{equation}
    l(\theta) = \sum_{j\in \mathcal{G} \cup \mathcal{G}_u} l_j(\theta_j)
\end{equation}

\noindent Here, $\theta_j$ denotes a matching for group $j$, and $\theta=\{\theta_j\}$ represents a global matching for a cluster. $\mathcal{G}$ and $\mathcal{G}_u$ represent the set of constrained and unconstrained groups in the cluster, respectively. Specifically, for each subset $s_i$ corresponding with group $j$, we can calculate $\max\limits_{\theta_j\in\Theta_j(s_i)}l_j(\theta_j)$ as the weight $w_{s_i}$ for this subset, as illustrated in the right-most column of Figure~\ref{fig:exact_cover}. Consequently, we can redefine the problem as a weighted version.

\begin{definition}[Weighted general exact cover problem]
Given a collection $S$ of subsets of a set $X$, where each subset $s \in S$ is associated with a weight $w_s \in \mathbb{R}$, the weighted general exact cover problem is to find a general exact cover $S^*$ that maximizes the total weight of the selected subsets.
\end{definition}

\begin{remark}
In the weighted general exact cover problem, we only care about whether selecting a subset affects the selection of other subsets, not the specific order of elements within this subset. Once the optimal solution is found, the matching $\theta_j$ with the highest log-likelihood function within $\Theta_j(s_i)$ is chosen as the final result for group $j$.
\end{remark}

\subsubsection{Solution algorithm}
Knuth's Algorithm X \citep{knuth_dancing_2000} is a backtracking algorithm that finds all solutions to a (general) exact cover problem. A technique named dancing links (DLX) is designed to implement Algorithm X efficiently, at which point the algorithm is also known as Algorithm DLX. The Algorithm DLX employs circular doubly linked lists for each row and column to store the 1s in the matrix instead of a two-dimensional array. Each column list also includes a special node known as the column header, which forms a special row consisting of all the columns that still exist in the matrix. This design accelerates operations such as removing and recovering columns or rows. However, since it is essentially a backtracking algorithm, Algorithm DLX has exponential complexity, $\mathcal{O}(c^n)$, where $c$ is a constant very close to 1, and $n$ is the number of $1$s in the matrix. In summary, the direct application of the Algorithm DLX faces the following challenges: (1) when the problem size increases, solution efficiency cannot be guaranteed, as a cluster in a real traffic scenario may contain dozens of vehicles; (2) it can only obtain all feasible solutions and cannot evaluate the quality of the solutions, hence it is unable to select the optimal solution.

To tackle these challenges, we have incorporated heuristics into the Algorithm DLX, as shown in Algorithm~\ref{algo:dlx}. The algorithm operates on a binary matrix $A$, where the objective is to find a general exact cover that maximizes the total weight, given a set of weights $w_r$ for each row $r$.

\paragraph{Matrix operation}
The function \texttt{Cover\_Row} is designed to cover all elements related to row $r$ in the matrix $A$. This is achieved by first removing all columns corresponding to the 1s in row $r$ from the dancing links structure and subsequently removing all rows corresponding to the 1s in those columns. In the context of our study, covering a row implies that if a possible matching for a group is selected, that group and its corresponding upstream vehicles are marked as used. Consequently, conflicting possible matchings are temporarily removed to prevent the reuse of upstream vehicles or the rematching of the group. Conversely, the function \texttt{Uncover\_Row} reverses the operations performed by \texttt{Cover\_Row}, restoring the matrix to its state prior to the covering. This reverse operation is crucial in the backtracking process.

\paragraph{Heuristic}
Our primary improvement to Algorithm DLX is the incorporation of heuristics to guide the search for an optimal solution. The \texttt{Heuristic} function estimates the potential weight contribution if row $r$ is included in the solution. It covers row $r$, computes an initial heuristic value as the weight of row $r$, and adds the maximum weight of the remaining rows for each primary column in the covered matrix. The matrix is then uncovered to preserve the original state.

\paragraph{Backtracking}
The \texttt{Backtracking} function is the core of our algorithm, which recursively explores potential solutions. If no primary columns are left in matrix $A$, the current solution and its weight are recorded as the best found so far. Otherwise, it selects a primary column with the fewest 1s, computes heuristics for the rows in this column, sorts them in descending order of their heuristic values, and iterates through each row. If the current solution's weight plus the heuristic does not exceed the best-known total weight, it prunes the search space. For each row, it covers the row, adds it to the current solution, and recurses. After exploring, it uncovers the row and backtracks.

\begin{remark}
When matrix $A$ is too large, consider the following measures to find a suboptimal solution:
\begin{itemize}
    \item Retain only the top $n$ rows (top $n$ most likely matchings) for each group by sorting the weights.
    \item Set a timeout for the algorithm.
\end{itemize}
\end{remark}

\begin{algorithm}[]
\caption{Algorithm DLX with heuristics}\label{algo:dlx}
\KwIn{The matrix $A$ consisting of 0s and 1s, weights $w_r$ for each row $r$}
\KwOut{Solution of the weighted general exact cover problem}
\SetKwProg{Function}{Function}{}{}
\SetKwFunction{Backtracking}{Backtracking}
\SetKwFunction{Heuristic}{Heuristic}
\SetKwFunction{CoverRow}{Cover\_Row}
\SetKwFunction{UncoverRow}{Uncover\_Row}

\BlankLine
\Function{\CoverRow{$A$, $r$}}{
    \ForEach{column $c$ such that $A[r,c]=1$}{
        Remove column $c$ from $A$\;
        \ForEach{row $i$ such that $A[i,c]=1$}{
            Remove row $i$ from $A$\;
        }
    }
}
\Function{\UncoverRow{$A$, $r$}}{
    \ForEach{column $c$ such that $A[r,c]=1$}{
        \ForEach{row $i$ such that $A[i,c]=1$}{
            Recover row $i$ from $A$\;
        }
        Recover column $c$ from $A$\;
    }
}
\BlankLine
\Function{\Heuristic{$A$, $r$}}{
    \CoverRow{$A$, $r$}\;
    $h \gets w_r$\;
    \ForEach{primary column $c'$ in matrix $A$}{
        $h \gets h + \max\limits_{r:A[r,c']=1} w_r $\;
    }
    \UncoverRow{$A$, $r$}\;
    \Return{$h$}
}
\BlankLine
\Function{\Backtracking{$A$, $cur\_sol$, $cur\_weight$}}{
    \If{matrix $A$ has no primary columns}{
        $sol \gets cur\_sol$ \;
        $total\_weight \gets cur\_weight$ \;
        \Return{}\;
    }
    Choose a primary column $c'$ with the lowest number of 1s\;
    $lst\_row \gets \{r : A[r,c']=1\}$\;
    \ForEach{row $r$ in $lst\_row$}{
        $h_r \gets$ \Heuristic{$A$, $r$}\;
    }
    Sort $lst\_row$ in descending order of $h_r$\;
    \ForEach{row $r$ in $lst\_row$}{
        \If{$total\_weight \neq -\infty$ and $cur\_weight + h_r \leq total\_weight$}{
            \Return{}\;
        }
        \CoverRow{$A$, $r$}\;
        Append $r$ to $cur\_sol$\;
        \Backtracking{$A$, $cur\_sol$, $cur\_weight + w_r$}\;
        Pop $r$ from $cur\_sol$\;
        \UncoverRow{$A$, $r$}\;
    }
}
\BlankLine
Initialize the optimal solution $sol$ to an empty list\;
Initialize the corresponding total weight $total\_weight$ to $-\infty$\;
\Backtracking{$A$, $[], 0$}\;
\Return{$sol$, $total\_weight$}\;
\end{algorithm}

\section{Evaluation}\label{sec:evaluation}

The single-lane estimation approach was first evaluated using an empirical case study. For comparison, the method proposed by \citet{zhan_lane-based_2015} using multi-section LPR data for lane-based queue length estimation, was also tested. This method is referred to as the Gaussian-process-car-following (GP-CF) method. Subsequently, the multi-lane estimation approach was evaluated using a simulation case study, accompanied by a sensitivity analysis of matching rates, volume-to-capacity (V/C) ratios, and FIFO violation rates.

\subsection{Calibration of the key parameters}
In the proposed approach, four key parameters require calibration: the minimum inter-group departure gap ($min\_gap$), the saturation headway ($h$), the discharging speed ($v_d$), and the delay threshold ($D_{thr}$). The minimum inter-group departure gap can be calibrated as the range of the running time. To calibrate the saturation headway, we employ the 15th percentile of observed headways on the target lane. Since the discharging speed and the delay threshold cannot be directly obtained from the LPR data, we utilize a grid search approach to systematically explore the parameter space and minimize the following loss function:

\begin{equation}
L(v_d, D_{thr}) = \alpha \sum_{c=1}^{N_{calib.}} \left(\hat{y}_c(v_d, D_{thr}) - y_c\right)^2 + \beta \sum_{c=1}^{N_{calib.}} \left(\hat{t}_{q,c}^{max}(v_d, D_{thr}) - t_{q,c}^{max}\right)^2
\end{equation}

\noindent where $N_{calib.}$ signifies the number of calibration cycles; $\hat{y}_c(v_d,D_{thr})$ and $\hat{t}_{q,c}^{max}(v_d, D_{thr})$ denote the estimated maximum queue length and the time of occurrence of the maximum queue length for cycle $c$ given parameters $v_d$ and $D_{thr}$, respectively; $y_c$ and $t_{q,c}^{max}$ represent the corresponding ground truth; while $\alpha$ and $\beta$ are weights assigned to the two terms. Importantly, $v_d$ and $D_{thr}$ only participate in the final step described in Section~\ref{sec:single_lane}, thereby facilitating a rapid search process. For the exclusive purpose of estimating the maximum queue length, calibration is required solely for the delay threshold by minimizing the loss function defined as

\begin{equation}
L(D_{thr}) = \sum_{c=1}^{N_{calib.}} \left(\hat{y}_c(v_d, D_{thr}) - y_c\right)^2
\end{equation}

\subsection{Empirical case study}
\begin{figure}[H]
    \centering
    \subfigure[Study site]{
        \includegraphics[width=0.45\textwidth]{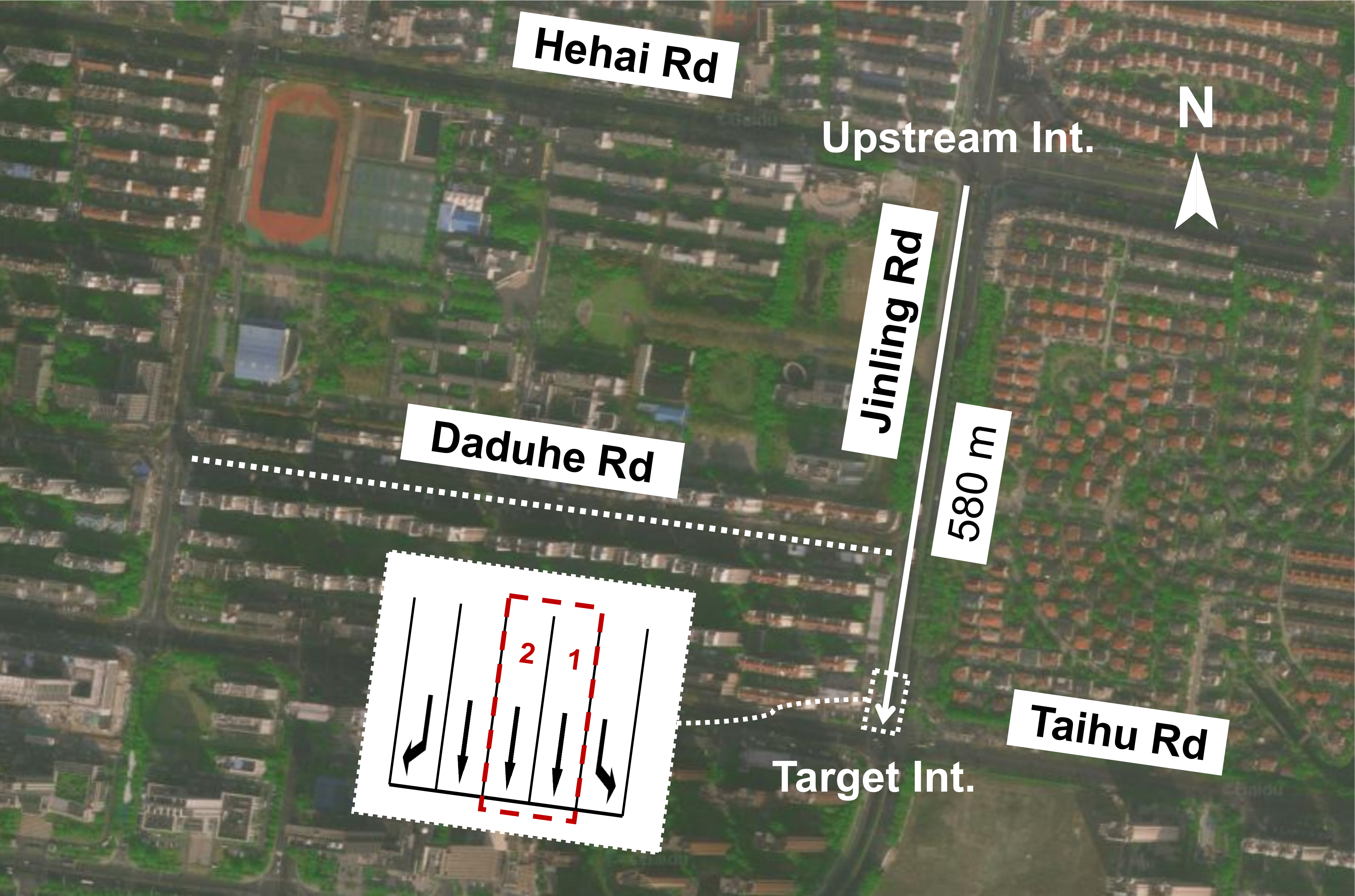}
        \label{fig:changzhou_site}
    }
    \hfill
    \subfigure[Signal phase and timing]{
        \includegraphics[width=0.45\textwidth]{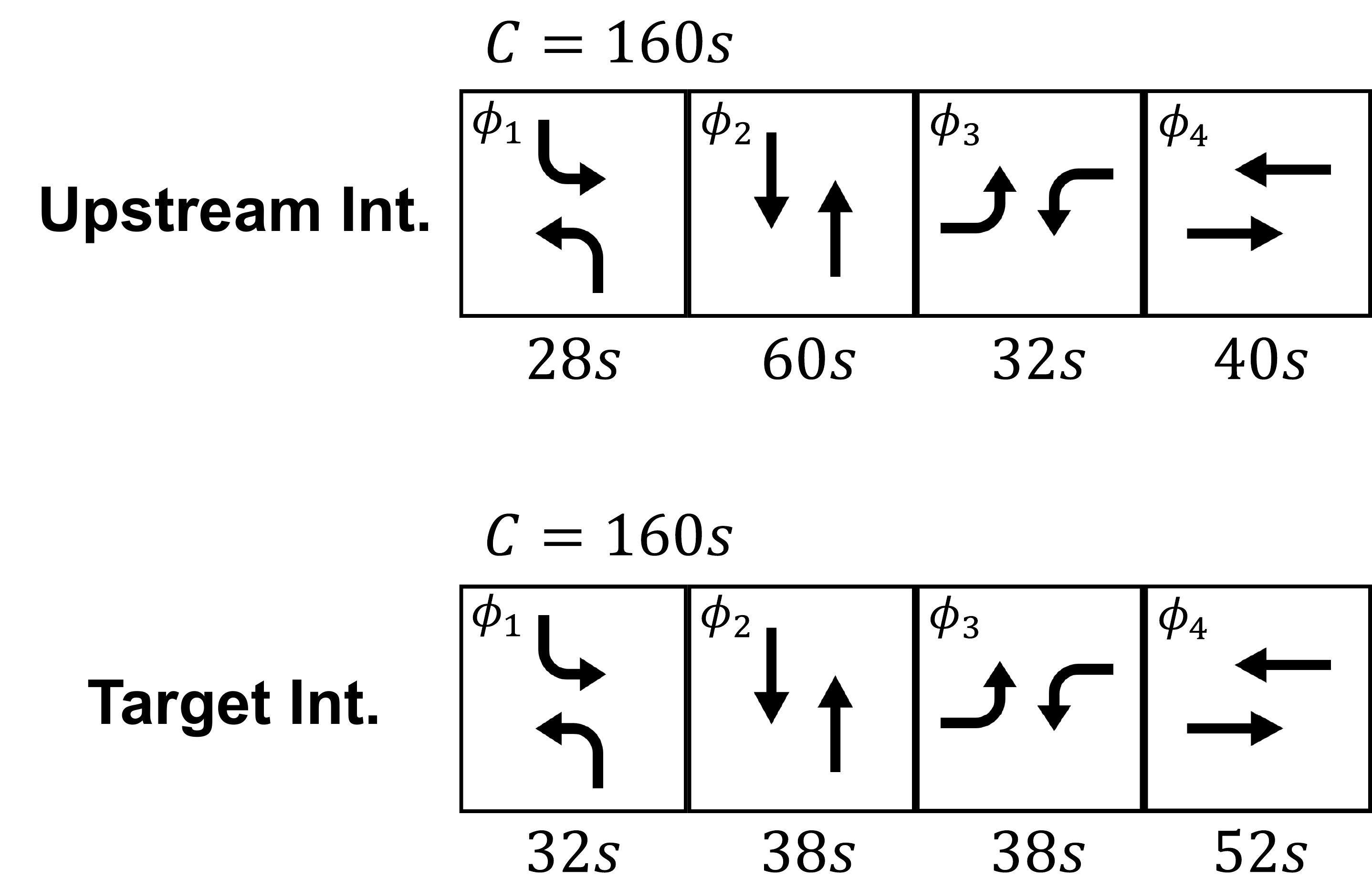}
        \label{fig:changzhou_timing}
    }
    \caption{Illustration of the empirical case study site.}
\end{figure}

The intersection of Jinling Road and Taihu Road in Changzhou, China, was selected for the empirical case study. Two through lanes of the southbound approach were studied, as illustrated in Figure \ref{fig:changzhou_site}. The upstream intersection is 580 meters away. All lanes at both intersections are captured by LPR cameras. Due to a branch road (Daduhe Road) within the link, vehicles in the studied lanes are not necessarily captured by the upstream intersection. Therefore, experiments for the multi-lane estimation were not conducted here. Given that branch roads within links are common in real-world scenarios, selecting this site helps to evaluate the general applicability of the proposed approach.

The LPR data and the corresponding recorded videos from both the upstream and the target intersections were available on September 7th, 2020. We selected the morning peak period from 7:00 am to 8:50 am as the study period. The cycle-by-cycle maximum queue lengths of the two studied lanes were obtained as the ground truth data using the opposite side camera. The recorded videos also captured the traffic light, so signal timing data could be extracted manually. Both the upstream and the target intersections adopted fixed timing plans with a command cycle length of 160 seconds, as illustrated in Figure \ref{fig:changzhou_timing}.

Subsequently, we proceed to the license plate matching process. The average matching rate is 87.5\%, which decreases to 73.0\% after filtering out FIFO violations for the GP-CF method. Video verification reveals a miss detection rate of less than 1\%. We utilize the ground truth queue length data from the first 8 cycles to calibrate the delay threshold $D_{thr}$ of the proposed approach and the car-following model of the GP-CF method. Considering the limited number of vehicles traveling at free-flow speed during peak hours, we utilize the travel time data from 6:00 am to 12:00 am to estimate the distribution of running time, thereby obtaining the minimum inter-group departure gap $min\_gap$. The calibrated values are summarized in Table~\ref{tab:parameters}.

\begin{table}
    \centering
    \caption{Parameters for the proposed approach.}
    \label{tab:parameters}
    \begin{tabular}{ccl}
        \toprule
        Parameter & Value & \multicolumn{1}{c}{Description}\\
        \midrule
        $h$ & 2.0 & Saturation headway (s)\\
        $D_{thr}$ & 5.1 & Delay threshold (s)\\
        $\mu$ & 3.89 & Location parameter of the log-normal distribution\\
        $\sigma$ & 0.15 & Scale parameter of the log-normal distribution\\
        $a$ & 31.1 & Minimum running time (s)\\
        $b$ & 58.0 & Maximum running time (s)\\
        $min\_gap$ & 26.9 & Minimum inter-group departure gap (s)\\
        \bottomrule
    \end{tabular}
\end{table}

Since the calculation result of the maximum queue length is in probabilistic form, to verify it, we first calculate the mean $\mu_{Q_c}$ of $Q_c$, and the lower $L_{Q_c}$ and upper $U_{Q_c}$ bounds of the 95\% confidence interval (CI):

\begin{gather}
    \mu_{Q_c} = \sum_i i \cdot P(Q_c = i)\\
    L_{Q_c} = \max_{P(Q_c \geq i) \geq 0.975} i\\
    U_{Q_c} = \min_{P(Q_c \geq i) \leq 0.025} i
\end{gather}

\noindent Then the performance of the maximum queue length estimation was evaluated using four metrics: the mean absolute error (MAE), the root mean square error (RMSE), the mean absolute percentage error (MAPE), and the coverage rate. These metrics were calculated as follows:

\begin{gather}
    MAE(\mathbf{y}, \hat{\mathbf{y}}) = \frac{1}{N_{cycle}} \sum_{c=1}^{N_{cycle}} |y_c - \hat{y}_c|\\
    RMSE(\mathbf{y}, \hat{\mathbf{y}}) = \sqrt{\frac{1}{N_{cycle}} \sum_{c=1}^{N_{cycle}} (y_c - \hat{y}_c)^2}\\
    MAPE(\mathbf{y}, \hat{\mathbf{y}}) = \frac{1}{N_{cycle}} \sum_{c=1}^{N_{cycle}} |\frac{y_c - \hat{y}_c}{y_c}| \times 100\%\\
    coverage\_rate = \frac{N_{covered}}{N_{cycle}} \times 100\%
\end{gather}

\noindent where $\mathbf{y}$ is the ground truth value; $\hat{\mathbf{y}}$ is the estimated value; $N_{cycle}$ is the number of cycles; and $N_{covered}$ is the number of cycles covered by the 95\% confidence interval.

\subsubsection{Maximum queue length estimation}

\begin{figure}[H]
    \centering
    \subfigure[Through lane 1]{
        \includegraphics[width=0.7\textwidth]{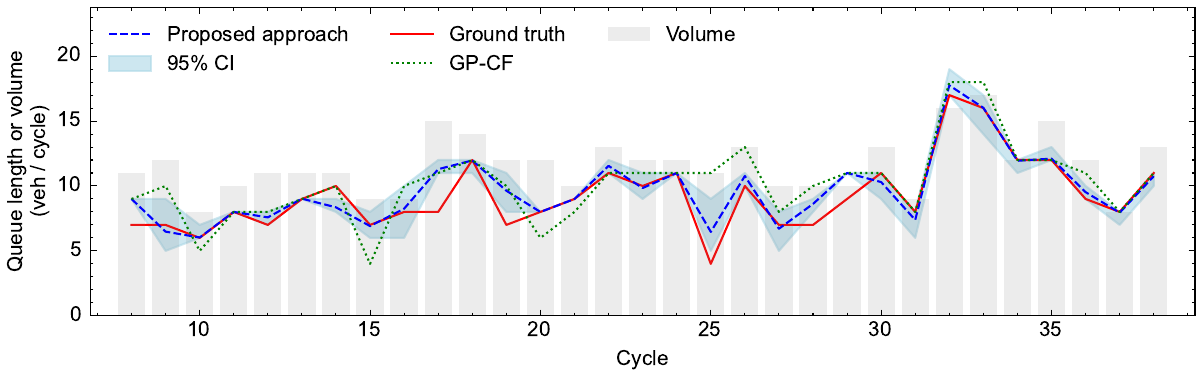}
        \label{fig:cz_queue_length_ds1}
    }
    \vfill
    \subfigure[Through lane 2]{
        \includegraphics[width=0.7\textwidth]{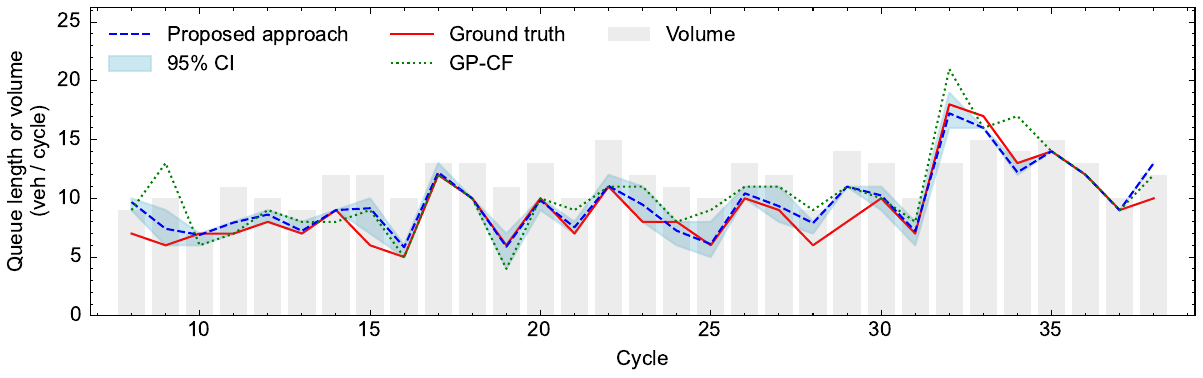}
        \label{fig:cz_queue_length_ds2}
    }
    \caption{Comparison of estimated maximum queue length.}
    \label{fig:cz_queue_length}
\end{figure}

\begin{figure}[H]
    \centering
    \subfigure[Proposed approach]{
        \includegraphics[width=0.4\textwidth]{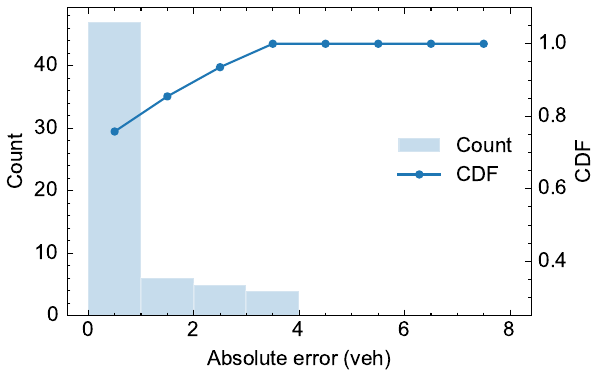}
        \label{fig:cz_error_hist_proposed}
    }
    \subfigure[GP-CF]{
        \includegraphics[width=0.4\textwidth]{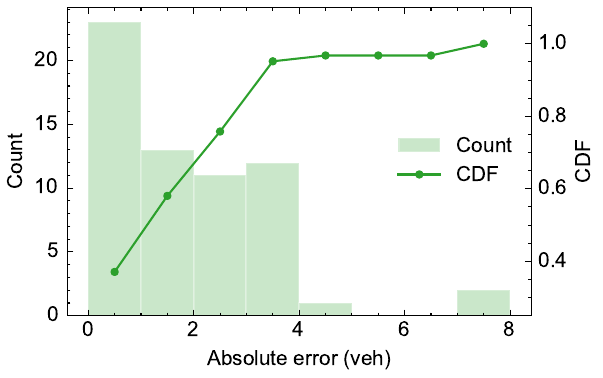}
        \label{fig:cz_error_hist_gp-cf}
    }
    \caption{Statistics of absolute errors in maximum queue length.}
    \label{fig:cz_error_hist}
\end{figure}

Figure~\ref{fig:cz_queue_length} presents the detailed queue length estimation results of the proposed approach and the GP-CF method, where the gray bars represent the traffic volume. The estimation error of the proposed approach is significantly smaller than the GP-CF method for half the cycles in both lanes, and the difference between the two is also relatively small in other cycles. From the error histograms in Figure~\ref{fig:cz_error_hist}, we can find that the majority of the errors for the proposed approach are concentrated below 1 veh, accounting for 76\% of all cycles of two lanes, with the maximum error being less than 4 veh. In contrast, errors below 1 veh only account for 37\% of the cycles for the GP-CF method, while most errors in other cycles are uniformly distributed between 1 veh and 4 veh, with a maximum error of 7 veh. These findings demonstrate that the proposed approach significantly outperforms the GP-CF method in terms of accuracy and reliability in empirical cases. Another advantage of the proposed approach is that we provide confidence intervals, and it can be found that the vast majority of cycles fall within the confidence intervals. Existing studies have demonstrated that such CI information can improve the performance of traffic signal control by robust theory \citep{tan_connected_2024}.

The overall performance of the two methods is summarized in Table~\ref{tab:estimation_results}. For all metrics, the improvement of the proposed approach exceeds 40\% compared to the GP-CF method. Furthermore, the coverage rate of the proposed approach also exceeds 75\% in both lanes.

\begin{table}[H]
    \centering
    \caption{Maximum queue length estimation results}
    \label{tab:estimation_results}
    \begin{tabular}{cccc}
        \toprule
        Metric & Proposed approach & GP-CF & Improvement \\ \midrule
        \multicolumn{4}{c}{\textit{Through lane 1}} \\ \midrule
        MAE (veh/cycle) & 0.71 & 1.39 & 48.92\% \\ 
        RMSE (veh/cycle) & 1.15 & 2.07 & 44.44\% \\ 
        MAPE (\%) & 9.65 & 20.05 & 51.87\% \\ 
        Coverage rate (\%) & 77.42 & N/A & N/A \\ \midrule
        \multicolumn{4}{c}{\textit{Through lane 2}} \\ \midrule
        MAE (veh/cycle) & 0.81 & 1.48 & 45.27\% \\ 
        RMSE (veh/cycle) & 1.25 & 2.17 & 42.40\% \\ 
        MAPE (\%) & 10.61 & 19.49 & 45.56\% \\ 
        Coverage rate (\%) & 80.65 & N/A & N/A \\ 
        \bottomrule
    \end{tabular}
\end{table}

\subsubsection{Impact of matching rates}
In the worst case, the matching rate of the LPR data at two consecutive intersections can degrade to 50-70\% \citep{zhan_lane-based_2015}. To test the performance of the proposed approach under extreme conditions, we intentionally degraded the matching rate by randomly removing vehicle IDs in the database using five different random seeds, resulting in six test scenarios with matching rates ranging from 30\% to 80\%.

\begin{figure}[H]
    \centering
    \subfigure[Through lane 1]{
        \includegraphics[width=0.9\textwidth]{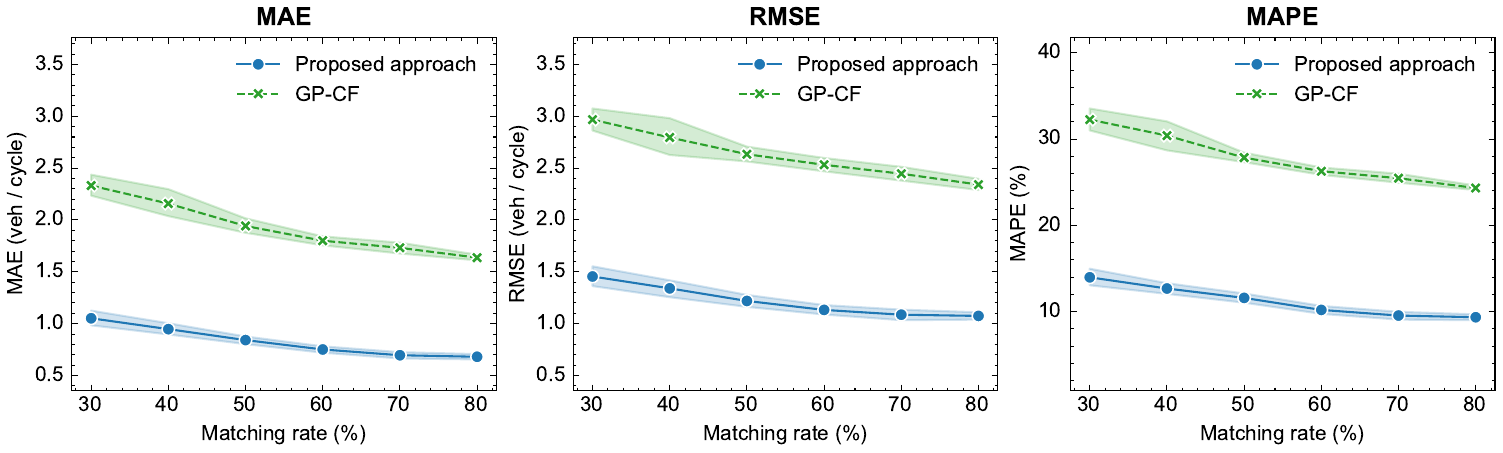}
        \label{fig:cz_queue_length_matching_rate_ds1}
    }
    \vfill
    \subfigure[Through lane 2]{
        \includegraphics[width=0.9\textwidth]{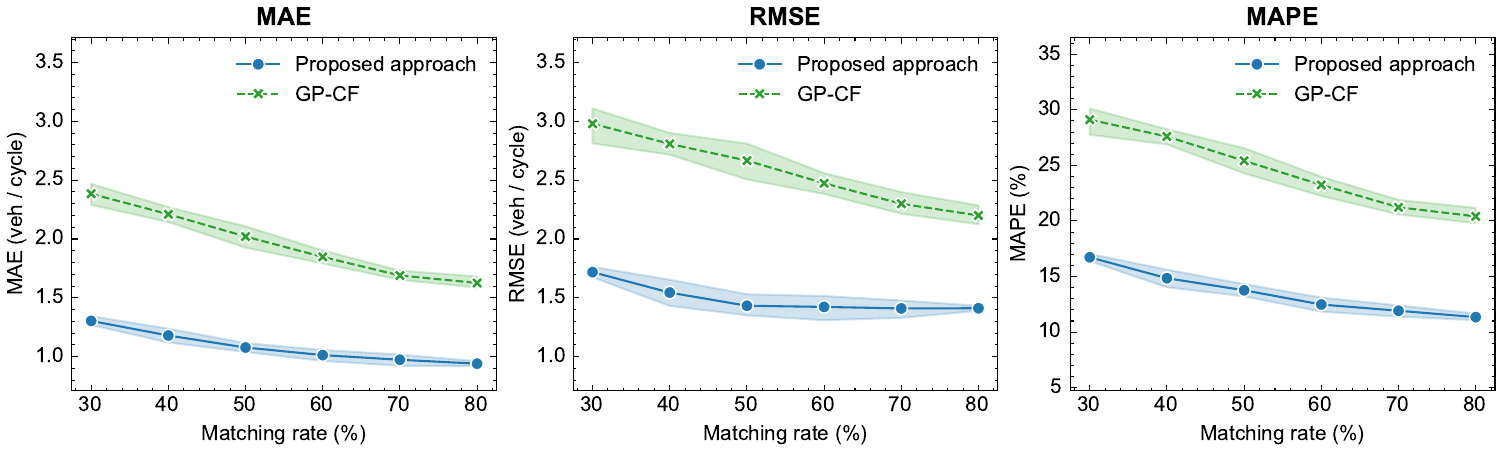}
        \label{fig:cz_queue_length_matching_rate_ds2}
    }
    \caption{Sensitivity analysis of maximum queue length estimation across various matching rates.}
    \label{fig:cz_queue_length_matching_rate}
\end{figure}

The results are shown in Figure~\ref{fig:cz_queue_length_matching_rate}. It can be observed that as the matching rate increases, the performance metrics of both methods improve. Although the improvement rate of both methods is similar as the matching rate increases, the proposed approach consistently outperforms the GP-CF method. Even at a matching rate of 30\%, the proposed approach outperforms the GP-CF method at a matching rate of 80\%. Additionally, unlike the GP-CF method, which shows a uniform decrease across all metrics, the proposed approach has a distinct inflection point. For example, the RMSE for through lane 2 decreases from 1.69 veh/cycle at a matching rate of 30\% to 1.41 veh/cycle at a matching rate of 50\%, then plateaus. Considering all metrics for both lanes, the matching rate of 60\% serves as a critical inflection point, beyond which further improvement in the matching rate yields diminishing returns in terms of estimation error reduction. This demonstrates that the proposed approach exhibits robustness to the matching rate and can still provide relatively accurate queue length estimates even in adverse detection environments.

\subsection{Simulation case study}
\begin{figure}[H]
    \centering
    \subfigure[Study site]{
        \includegraphics[width=0.45\textwidth]{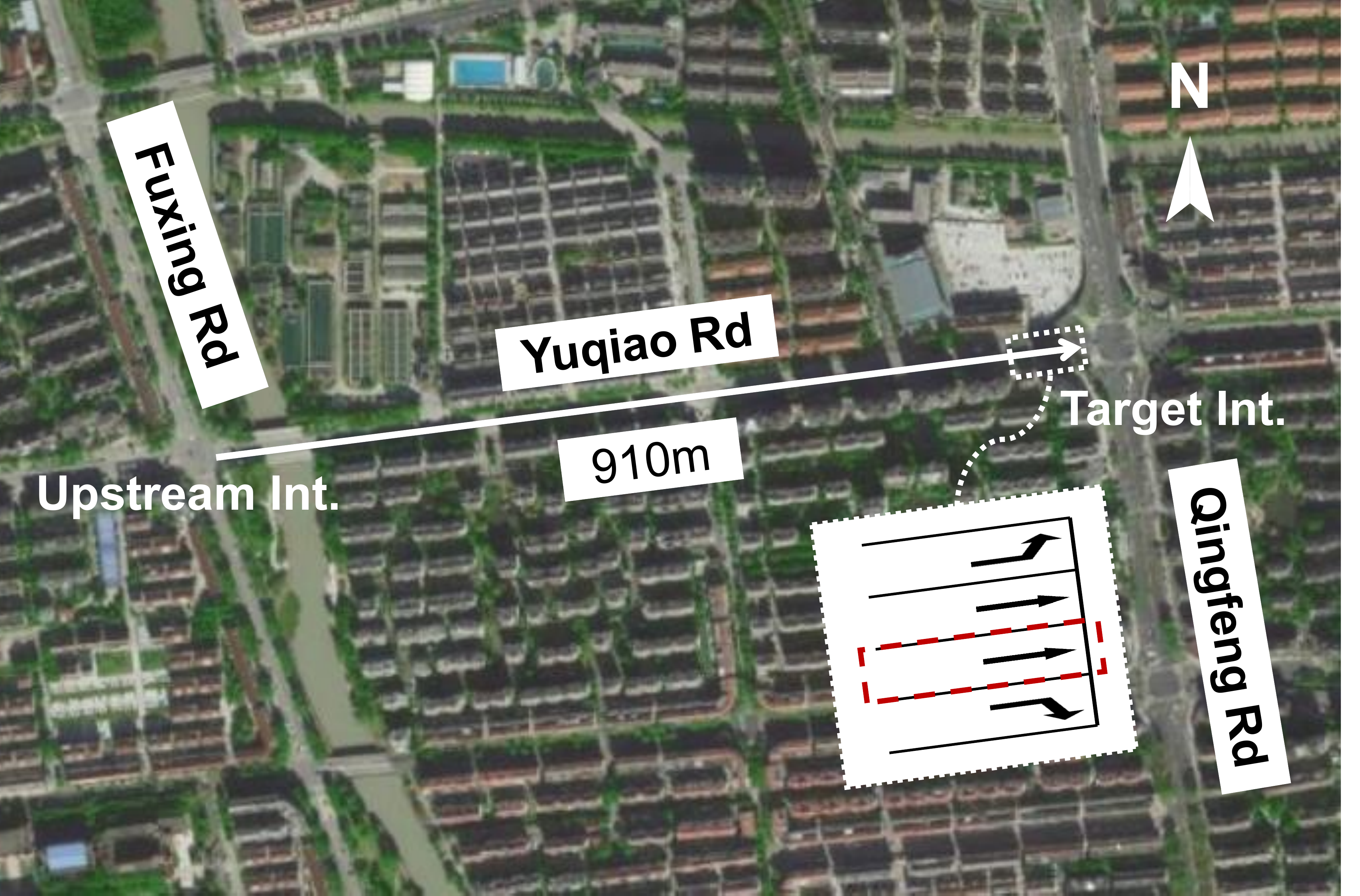}
        \label{fig:simulation_site}
    }
    \hfill
    \subfigure[Signal phase and timing]{
        \includegraphics[width=0.45\textwidth]{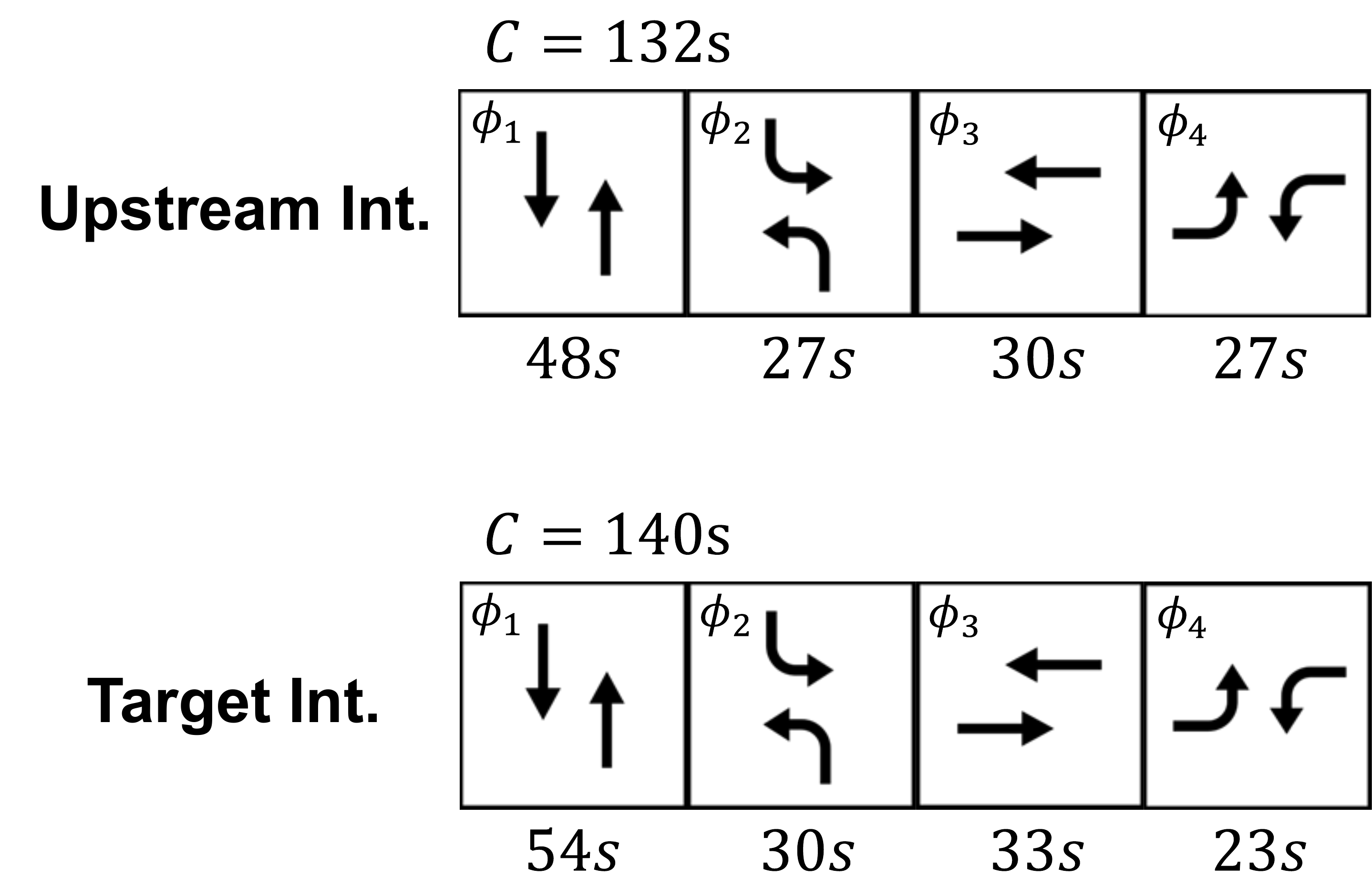}
        \label{fig:simulation_timing}
    }
    \caption{Illustration of the simulation case study site.}
\end{figure}

To evaluate the multi-lane estimation approach, we constructed a small network comprising two intersections in Tongxiang, China, and implemented the simulation on the SUMO platform. As shown in Figure~\ref{fig:simulation_site}, one through lane of the eastbound approach at the intersection of Qingfeng Road and Yuqiao Road was selected for study. The stop line of the approach is about 910 meters from the upstream, and variations in travel times and overtaking behaviors are significantly different from the empirical case. The signal phases and timing are illustrated in Figure~\ref{fig:simulation_timing}, showing that the two intersections are not coordinated. The simulation model was calibrated using LPR data from 8:00 am to 9:00 am, including arrival rate, speed, turning ratio, and saturation headway. Then, based on the basic scenario, we considered different V/C ratios, matching rates, and FIFO violation rates, which yielded various simulation scenarios. The simulation ran for 7800 seconds with a warm-up period of 600 seconds. The trajectory data of each vehicle, including vehicle IDs, timestamps, and locations, were recorded in the database to verify queue length estimation. Finally, we used the first 12 cycles after the warm-up period to calibrate the proposed approach for each scenario, including the discharging speed $v_d$ and the delay threshold $D_{thr}$. Both weights $\alpha$ and $\beta$ were assigned a value of 1. The car-following models of the GP-CF method were calibrated accordingly.

Since we can obtain the true queue profile in the simulation, the estimated queue profile is evaluated in this section. Like the evaluation of the maximum queue length, we first calculate the mean $\mu_{Q(t)}$ of $Q(t)$, and the lower $L_{Q(t)}$ and upper $U_{Q(t)}$ bounds of the 95\% CI:

\begin{gather}
    \mu_{Q(t)} = \sum_i i \cdot P(Q(t) = i)\\
    L_{Q(t)} = \max_{P\left(Q(t) \geq i\right) \geq 0.975} i\\
    U_{Q(t)} = \min_{P\left(Q(t) \geq i\right) \leq 0.025} i
\end{gather}

\noindent Then our queue profile estimation performance was evaluated using three metrics: the MAE during red, the MAE during green, and the coverage rate. These metrics were calculated as follows:

\begin{gather}
    MAE_{red}(\mathbf{y}, \hat{\mathbf{y}}) = \frac{1}{|\mathcal{T}_{red}|}\sum_{t\in \mathcal{T}_{red}}|y_t-\hat{y_t}|\\
    MAE_{green}(\mathbf{y}, \hat{\mathbf{y}}) = \frac{1}{|\mathcal{T}_{green}|}\sum_{t\in \mathcal{T}_{green}}|y_t-\hat{y_t}|\\
    coverage\_rate = \frac{|\mathcal{T}_{covered}|}{|\mathcal{T}|} \times 100\%
\end{gather}

\noindent where $\mathbf{y}$ is the ground truth value; $\hat{\mathbf{y}}$ is the estimated value; $\mathcal{T}=\mathcal{T}_{red}\cup\mathcal{T}_{green}$ is the set of time steps in the study period; $\mathcal{T}_{red}$ is the set of time steps during red phases; $\mathcal{T}_{green}$ is the set of time steps during yellow and green phases; $\mathcal{T}_{covered}$ is the set of time steps covered by the 95\% confidence interval. The reason for separately calculating the error during green phases is that the queue length will instantly drop to zero at some point during the green phase, causing small deviations to result in a large MAE, and part of the green phase has no queue.

\subsubsection{Impact of V/C ratios}
In this section, we scaled the arrival rate of the upstream intersection to create five scenarios with varying V/C ratios, ranging from 0.5 to 0.9. We then randomly removed vehicle IDs from the database using five different random seeds, obtaining results with a matching rate of 60\% as representative.

\begin{figure}[H]
    \centering
    \subfigure[Maximum queue length estimation]{
        \includegraphics[width=1\textwidth]{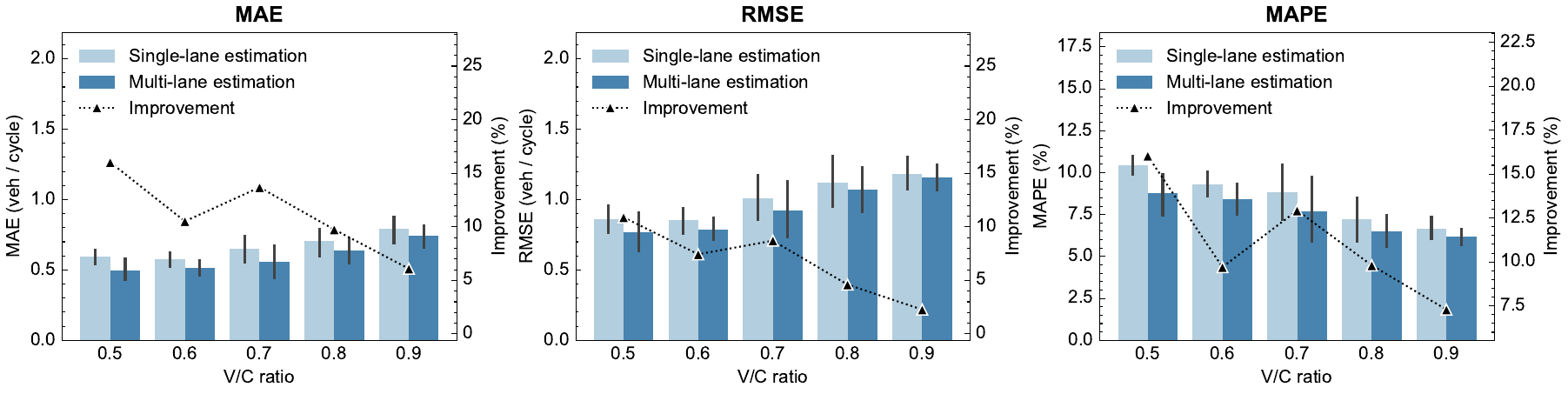}
        \label{fig:queue_length_vc_ratio}
    }
    \vfill
    \subfigure[Queue profile estimation]{
        \includegraphics[width=1\textwidth]{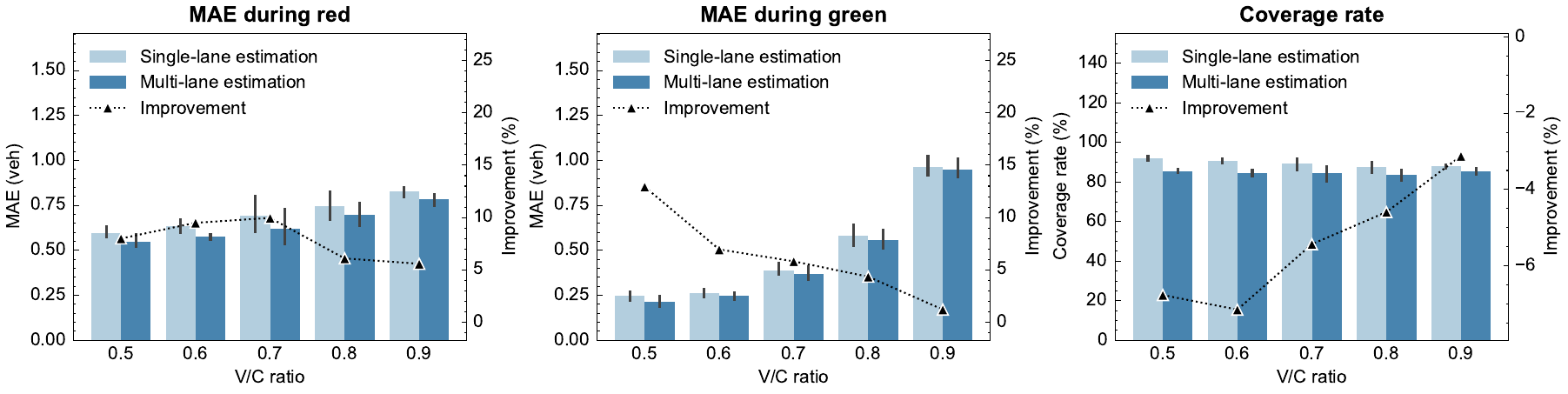}
        \label{fig:queue_profile_vc_ratio}
    }
    \caption{Sensitivity analysis of queue length estimation across various V/C ratios.}
    \label{fig:sensitivity_vc_ratio}
\end{figure}

\textbf{Maximum queue length estimation:}
As illustrated in Figure~\ref{fig:queue_length_vc_ratio}, the multi-lane estimation consistently outperforms the single-lane estimation across all V/C ratio scenarios. With increasing V/C ratios, the MAE and RMSE for both estimation approaches increase, peaking at a V/C ratio of 0.9, consistently staying below 1.0 veh/cycle and 1.5 veh/cycle, respectively. However, the MAPE decreases as the V/C ratio rises due to the larger ground truth in high V/C ratio scenarios, and remains below 12\%. The improvement percentage of the multi-lane estimation over the single-lane estimation is significant at lower V/C ratios (0.5 and 0.7) but diminishes as the V/C ratio increases.

\textbf{Queue profile estimation:}
Figure~\ref{fig:queue_profile_vc_ratio} shows that the multi-lane estimation remains superior to the single-lane estimation in all scenarios. The MAE during the red and green phases rises with the V/C ratio, with a more gradual increase during the red phase and a more substantial rise during the green phase. This trend is attributed to longer queues during the green phase in high V/C ratio scenarios, where minor deviations in queue clearance time can cause significant errors in the queue profile. The coverage rate is less affected by the V/C ratio, with a slight decrease in the single-lane estimation and relatively stable in the multi-lane estimation, both remaining above 80\%.

\textbf{Overall result:}
The proposed approach delivers robust results under various V/C ratios. The multi-lane estimation achieves better accuracy, particularly at medium V/C ratios (0.5 to 0.7).

\subsubsection{Impact of matching rates}
In the empirical case study, we tested the performance of the proposed single-lane estimation approach and the GP-CF method under varying matching rates, finding that a matching rate of 60\% serves as an inflection point, beyond which further improvements are marginal. In the simulation case study, we focused on comparing the performance of the single-lane and multi-lane estimations with a fixed V/C ratio of 0.7.

\begin{figure}[H]
    \centering
    \subfigure[Maximum queue length estimation]{
        \includegraphics[width=0.9\textwidth]{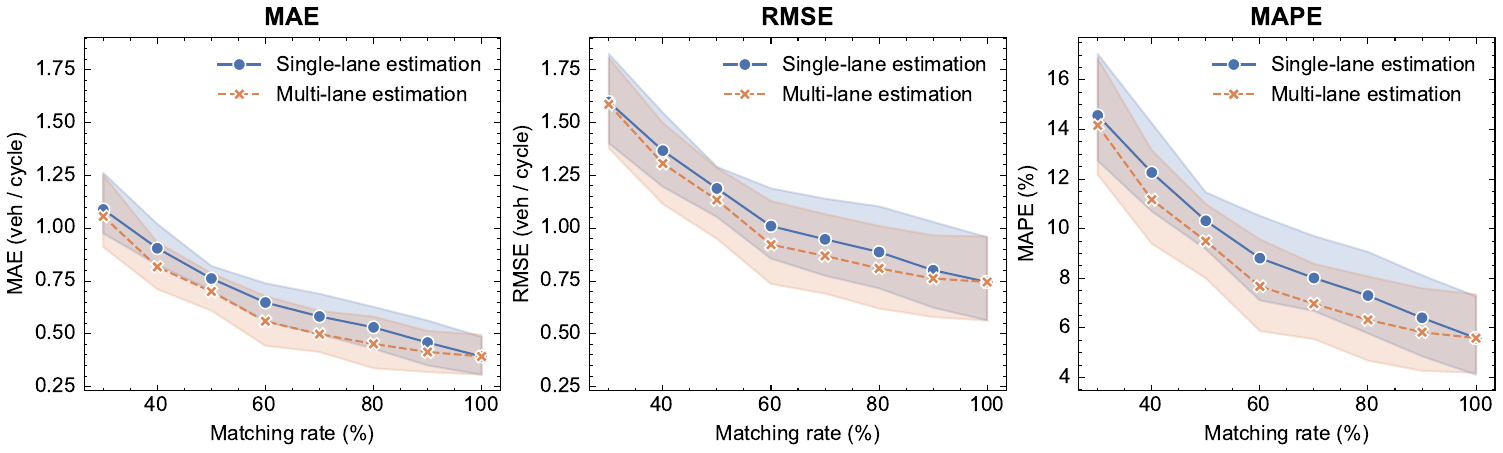}
        \label{fig:queue_length_matching_rate}
    }
    \vfill
    \subfigure[Queue profile estimation]{
        \includegraphics[width=0.9\textwidth]{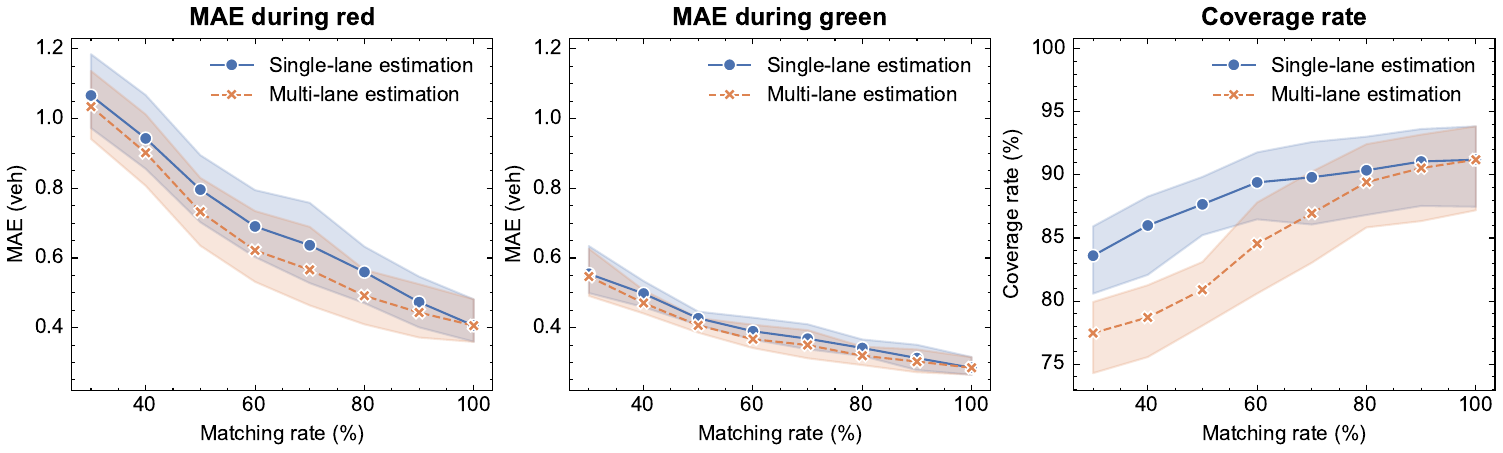}
        \label{fig:queue_profile_matching_rate}
    }
    \caption{Sensitivity analysis of queue length estimation across various matching rates.}
\end{figure}

\textbf{Maximum queue length estimation:}
As depicted in Figure~\ref{fig:queue_length_matching_rate}, a 60\% matching rate serves as an elbow point in the curve (more apparent in RMSE and MAPE), but beyond 60\%, each metric continues to decline. When the matching rate is between 40\% and 90\%, the multi-lane estimation shows certain improvements over the single-lane estimation, particularly notable at matching rates of 60\% to 80\%.

\textbf{Queue profile estimation:}
As depicted in Figure~\ref{fig:queue_profile_matching_rate}, both the single-lane and multi-lane estimations improve with higher matching rates. The MAE during red phases shows significant improvement in the multi-lane estimation, especially for matching rates between 50\% and 80\%, while improvements during green phases are less pronounced. The coverage rate increases with higher matching rates, eventually converging around 91\%. When the matching rate is less than 80\%, the coverage rate of the multi-lane estimation is significantly lower than that of the single-lane estimation. This trend is due to our efforts to reduce uncertainty by matching all unmatched vehicles in the multi-lane estimation. Even slight estimation errors can cause the ground truth to fall outside the 95\% CI.

\textbf{Overall result:}
The proposed approach is robust and effective in estimating maximum queue lengths and queue profiles under varying matching rates, with a matching rate of 60\% serving as an inflection point. Moreover, the multi-lane estimation shows significant improvements at most matching rates.

\subsubsection{Impact of FIFO violation rates}
\begin{figure}[H]
    \centering
    \includegraphics[width=0.4\textwidth]{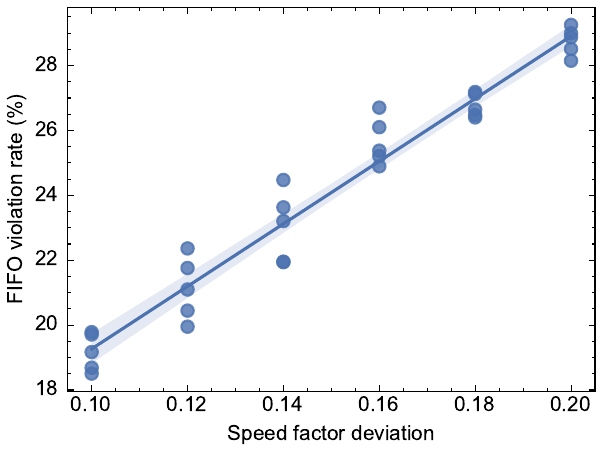}
    \caption{The correlation between the speed factor deviation and the FIFO violation rate.}
    \label{fig:speed_fifo}
\end{figure}

A key difference between the proposed approach and existing methods lies in the relaxation of the FIFO assumption, making it necessary to study the impact of the FIFO violation rate on this approach. To achieve this, we selected the scenario with a V/C ratio of 0.7 and a matching rate of 60\% and controlled the frequency of overtaking by adjusting the distribution of the speed factor in SUMO. The speed factor in SUMO is a vehicle-specific multiplier that, when applied to the road speed limit, determines a vehicle's desired free-flow driving speed. The default speed factor for passenger cars follows a truncated normal distribution $\mathcal{N}(1, 0.1, 0.2, 2)$, implying that about 95\% of the vehicles drive between 80\% and 120\% of the legal speed limit. Therefore, by increasing the deviation of the speed factor from 0.10 to 0.20, the change in the FIFO violation rate is shown in Figure~\ref{fig:speed_fifo}, revealing a clear linear relationship between the two. When the speed factor deviation is 0.1, approximately 19\% of vehicles violate FIFO, and when the speed factor deviation reaches 0.2, this proportion increases to 29\%.

\begin{figure}[H]
    \centering
    \subfigure[Maximum queue length estimation]{
        \includegraphics[width=1\textwidth]{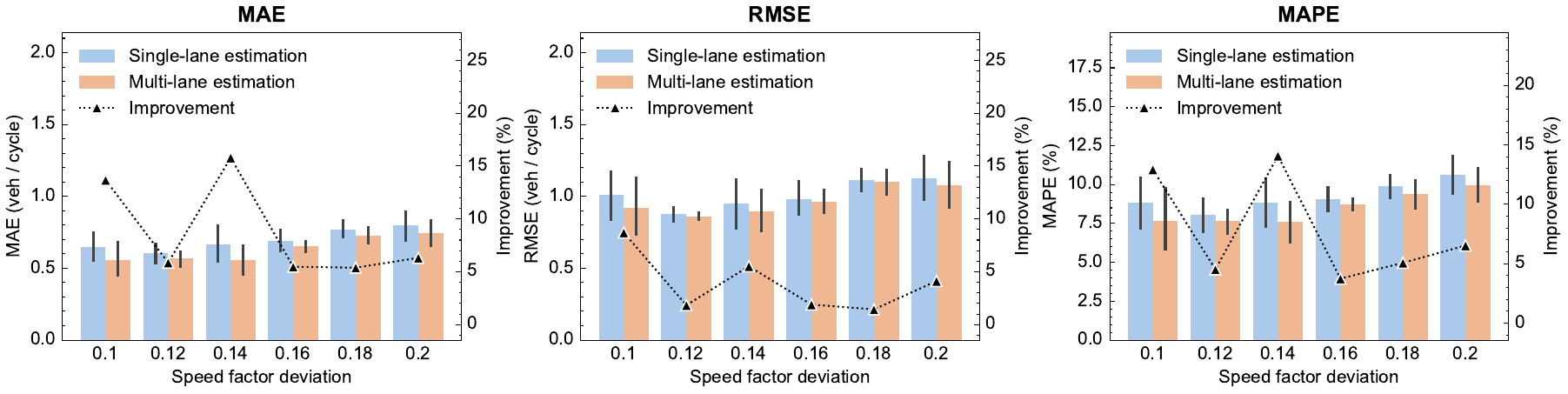}
        \label{fig:queue_length_speed_deviation}
    }
    \vfill
    \subfigure[Queue profile estimation]{
        \includegraphics[width=1\textwidth]{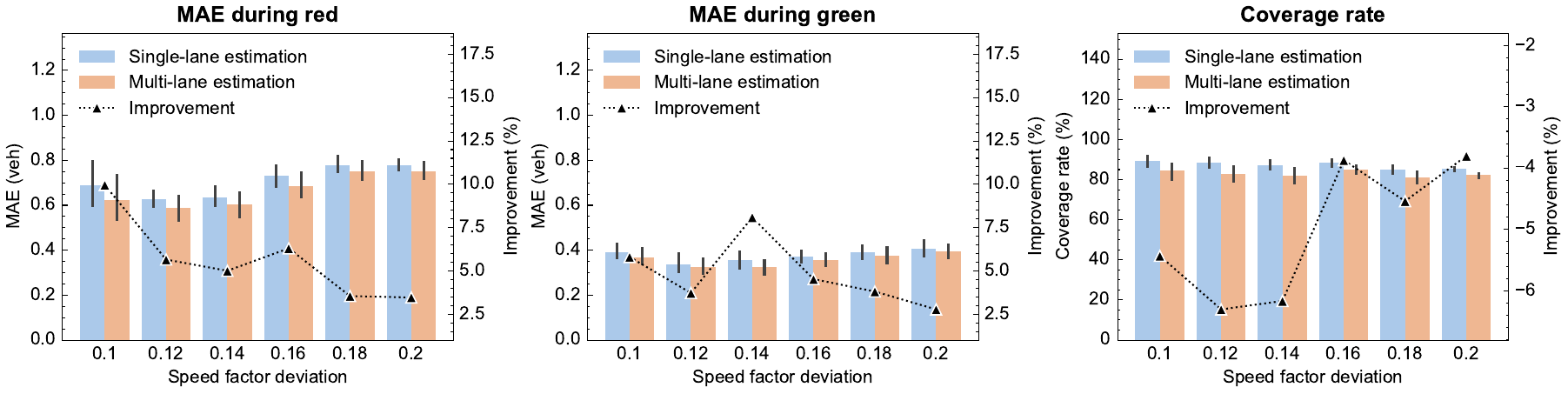}
        \label{fig:queue_profile_speed_deviation}
    }
    \caption{Sensitivity analysis of queue length estimation across various speed factor deviations.}
\end{figure}

\textbf{Maximum queue length estimation:}
The results illustrated in Figure~\ref{fig:queue_length_speed_deviation} show a slightly increasing trend in the metrics as the speed factor deviation increases. The average values of MAE, RMSE, and MAPE for both the single-lane and multi-lane estimations remain around 0.8, 1.2, and 11\%, respectively. The multi-lane estimation consistently outperforms the single-lane estimation, particularly at speed factor deviations of 0.1 and 0.14.

\textbf{Queue profile estimation:}
Similar trends are observed in Figure~\ref{fig:queue_profile_speed_deviation}. The average MAE values during the red and green phases are below 0.8 and 0.4, respectively. The coverage rate decreases slightly as the speed factor deviation increases but remains above 80\% for both the single-lane and multi-lane estimations.

\textbf{Overall result:}
The proposed approach exhibits robustness in the presence of frequent overtaking, even under extreme conditions. Specifically, with a speed factor deviation of 0.2, the performance remains stable, with 95\% of vehicles' expected speeds falling within 60\% to 140\% of the legal speed limit. This robustness can be attributed to our approach's greater emphasis on the departure sequence of the target lanes, rather than the upstream departure sequence, thereby considering the overtaking and the potential impact on vehicle running times.

\subsubsection{Detailed queue profile estimation}
\begin{figure}[H]
    \centering
    \includegraphics[width=0.9\textwidth]{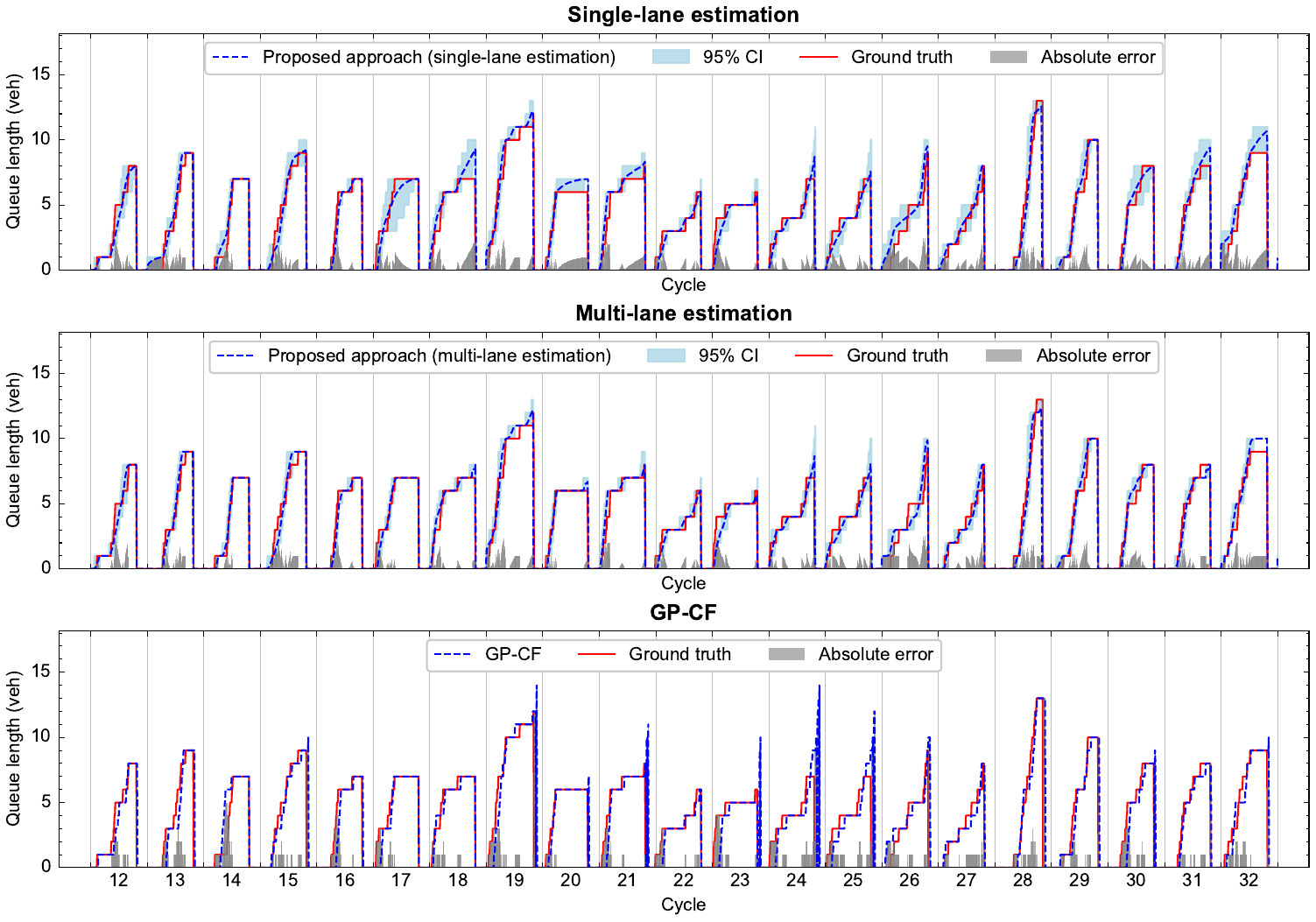}
    \caption{Partial cycle result of estimated queue profile with a V/C ratio of 0.7.}
    \label{fig:queue_profile_0.7}
\end{figure}

\begin{figure}[H]
    \centering
    \includegraphics[width=0.9\textwidth]{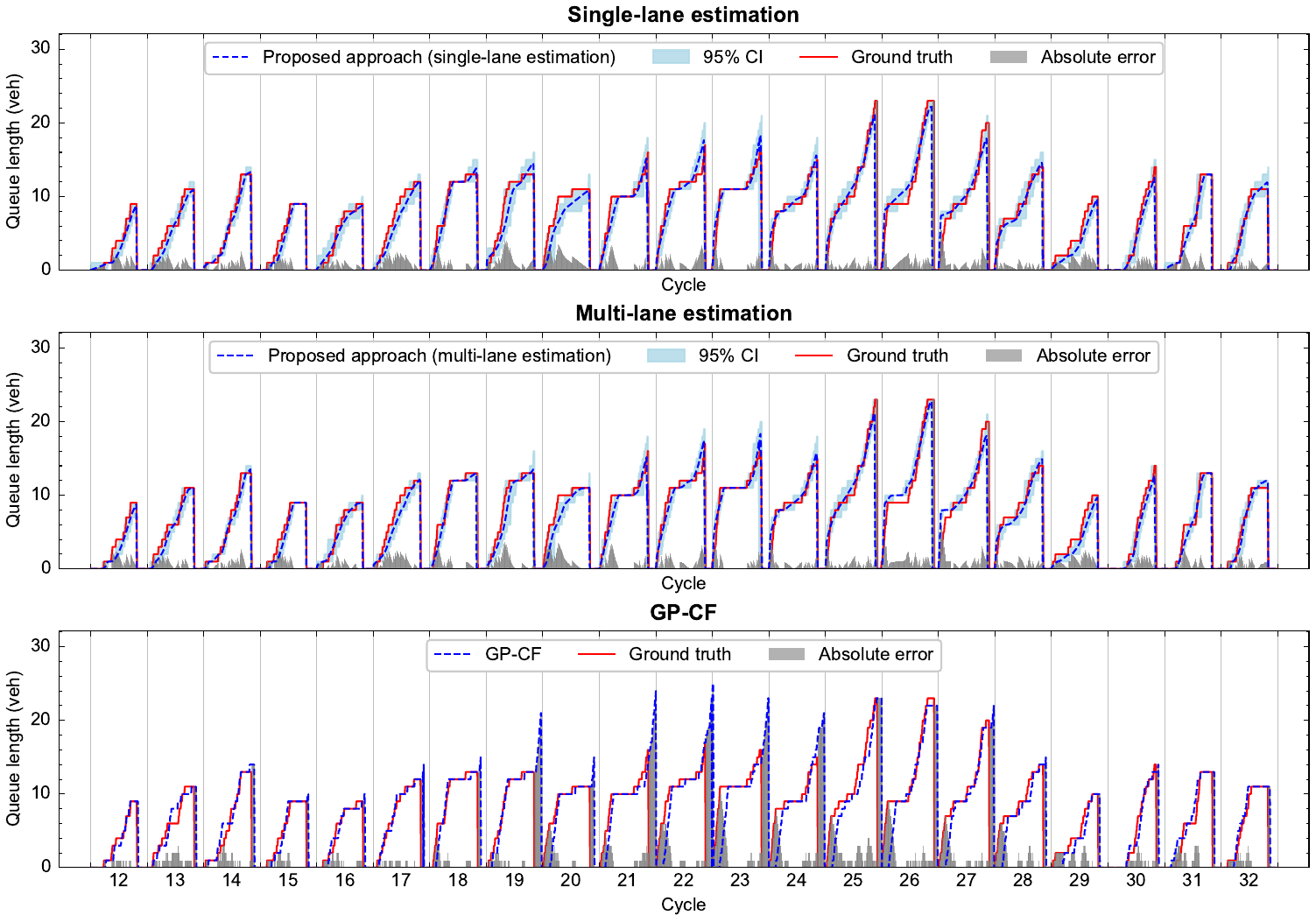}
    \caption{Partial cycle result of estimated queue profile with a V/C ratio of 0.9.}
    \label{fig:queue_profile_0.9}
\end{figure}

Figure~\ref{fig:queue_profile_0.7} and Figure~\ref{fig:queue_profile_0.9} present the partial cycle results of the estimated queue profiles from the proposed approach (including both the single-lane and multi-lane estimations) and the GP-CF method. Figure~\ref{fig:queue_profile_0.7} shows the results for a V/C ratio of 0.7, while Figure~\ref{fig:queue_profile_0.9} shows the results for a V/C ratio of 0.9, both with a matching rate of 60\%.

\textbf{V/C ratio of 0.7:} The proposed approach outperforms the GP-CF method, with the multi-lane estimation showing a slight advantage over the single-lane estimation. Since the queue mostly forms during the red phase, the improvement in the MAE during the red phase is more significant. For instance, in cycles 17, 18, 20, and 21, the multi-lane estimation results show noticeably smaller deviations from the ground truth, with a narrower 95\% CI, which also reduces the error in estimating the maximum queue length. The GP-CF method estimates the arrival curve using signal timing information from the upstream intersection, leading to a good estimation of the approximate upstream departure times and acceptable queue profile estimation during the queue formation period. However, since the GP-CF method relies on a car-following model that assumes only vehicles heading to the target lane are present, it is challenging to accurately determine the actual back of the queue.

\textbf{V/C ratio of 0.9:} Similar to the V/C ratio of 0.7, the proposed approach outperforms the GP-CF method across all metrics, though the difference between the single-lane and multi-lane estimations is smaller. Notably, some cycles in this scenario experience over-saturation, resulting in residual queues, such as cycles 22 to 28. For each over-saturated cycle, the GP-CF method tends to overestimate the maximum queue length, leading to a rightward shift in the queue profile at the start of the next cycle. In contrast, both the single-lane and multi-lane estimations accurately estimate the queue profile for these cycles, indicating the approach's suitability for over-saturated scenarios.

\begin{figure}[H]
    \centering
    \subfigure[V/C ratio of 0.7]{
        \includegraphics[width=0.9\textwidth]{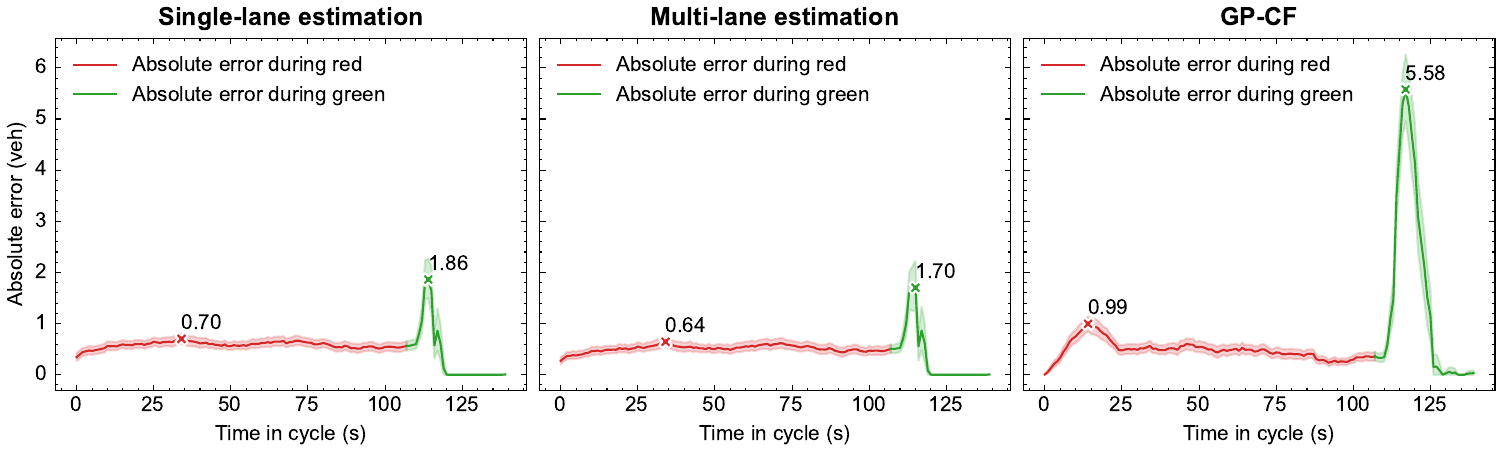}
        \label{fig:queue_profile_error_0.7}
    }
    \vfill
    \subfigure[V/C ratio of 0.9]{
        \includegraphics[width=0.9\textwidth]{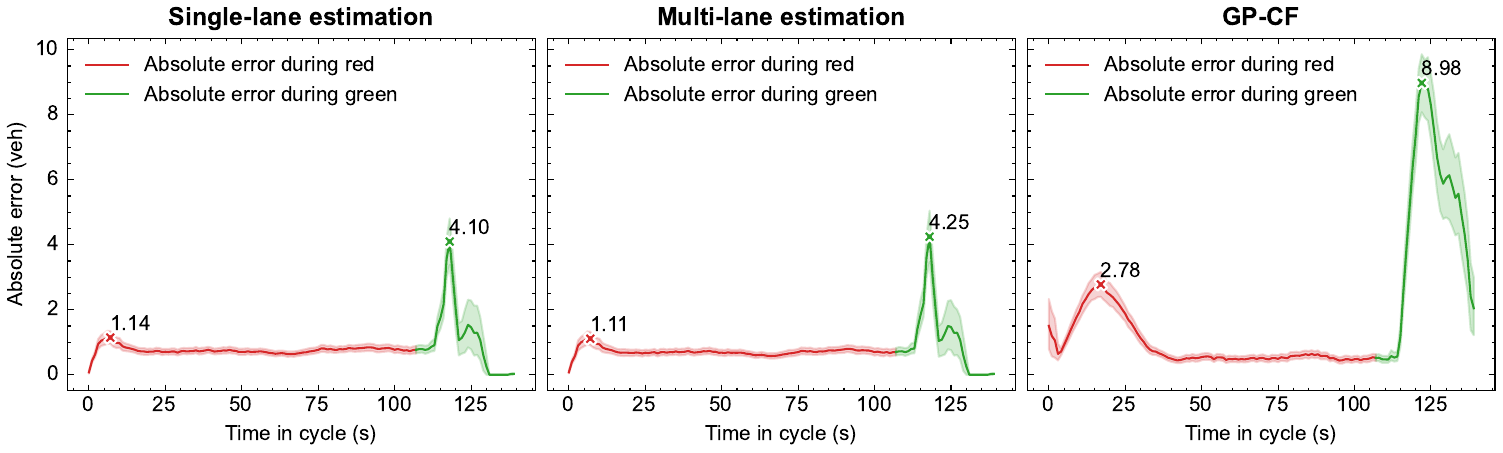}
        \label{fig:queue_profile_error_0.9}
    }
    \caption{Comparison of average errors of queue profile estimation with V/C ratios of 0.7 and 0.9.}
    \label{fig:queue_profile_error}
\end{figure}

\textbf{Error analysis:} To gain a deeper understanding of the estimation results, we averaged the absolute errors over multiple cycles within the study period, as illustrated in Figure~\ref{fig:queue_profile_error}. As shown, the absolute error during one cycle in both the single-lane and multi-lane estimations follows a similar trend. During the red phase, the absolute error continuously increases from its initial value, reaching a peak and then slightly decreasing. The absolute error increases more rapidly at a V/C ratio of 0.9, primarily due to the residual queue from over-saturated cycles. With the onset of the green light, the absolute error rapidly increases to the maximum value of the entire cycle and then quickly decreases to zero. The GP-CF method shows a slightly different pattern, with the absolute error rising faster during the red phase and peaking much higher during the green phase compared to the proposed approach. This is primarily due to the deviation at the moment when the queue fully clears. Notably, for the scenario with a V/C ratio of 0.9, the GP-CF method's curve shifts significantly to the right, which is consistent with the previously observed rightward shift in queue profile estimation results for over-saturated cycles.

\section{Conclusion}\label{sec:conclusion}

In this research, we presented a probabilistic approach for estimating lane-based queue length (including maximum queue length and queue profile) using multi-section LPR data. This approach relaxes the assumptions of the FIFO rule and specific arrival processes employed in previous studies, allowing for overtaking behaviors and application to multi-lane scenarios. We achieved this by introducing the concept of vehicles' NATs and focusing on estimating its distribution. First, we established NAT conditions for each vehicle group in two distinct categories: constrained vehicles and unconstrained vehicles, partitioned using a DP-based approach. Such a partitioning approach can significantly reduce the computational dimensionality and complexity. We subsequently adopted an MCMC sampling method to estimate the arrival distribution based on the prior running time distribution. By maximizing the number of groups, the vehicle count in each group reduces, hence enhancing sampling efficiency. After determining the arrival distribution of all vehicles, the distribution of the maximum queue length and the queue profile can be derived. To maximize the exploitation of upstream LPR data, we further extended the methodology from the single-lane estimation to the multi-lane estimation. We transformed the problem of finding the optimal matching of unmatched vehicles between the upstream and the target intersection into a weighted general exact cover problem, which was solved by the Algorithm DLX with heuristics. Subsequently, the arrival distribution of each group would be updated, facilitating more accurate queue length estimation.

The proposed approach was evaluated using both empirical and simulation case studies. The empirical case study demonstrated significant improvements over a state-of-the-art method, with over 40\% enhancements in MAE, RMSE, and MAPE for maximum queue length estimation. The simulation case study further validated the robustness of the proposed approach under various V/C ratios and FIFO violation rates. A matching rate of 60\% serves as an inflection point, showing the approach's applicability in diverse detection conditions. The multi-lane estimation outperforms the single-lane estimation in most scenarios. Detailed queue profile results confirm the approach's ability to handle over-saturated traffic conditions.

There are several possible directions for future research: (1) The proposed approach currently does not factor in the effects of short lanes. In future work, we could specifically model these lanes and devise an algorithm to estimate the queue spillback probability. (2) As LPR data cannot directly provide vehicle trajectories within links, exploring data fusion methods, such as integrating CV data \citep{tan_fuzing_2020}, can improve the accuracy of queue profile estimation. Furthermore, investigating algorithms for vehicle trajectory reconstruction based on LPR data or the fusion of LPR and CV data, utilizing car-following and lane-changing models, is also a promising direction.

\section*{Acknowledgment}
This research was sponsored by the National Key Research \& Development Program (Grant No. 2023YFB4301900) and the National Natural Science Foundation of China (Grant No. 52372319).

\begin{appendices}
\section{Log-likelihood function of global match $\theta$}\label{sec:appendixA}
Let $\theta_j$ denote a matching for group $j$, and let $\theta=\{\theta_j | j \in \mathcal{G} \cup \mathcal{G}_u\}$ denote a global matching for a cluster, where $\mathcal{G}$ and $\mathcal{G}_u$ represent the set of constrained and unconstrained groups in the cluster, respectively. Therefore, using the NAT conditions derived from Section~\ref{sec:NAT-conditions}, we can formulate a maximum likelihood estimation:

\begin{equation}
    \hat{\theta}=\underset{\theta \in \Theta}{\arg \max}\,
    P\left(\bigcap_{j \in \mathcal{G}} \mathcal{C}_j,\bigcap_{j \in \mathcal{G}_u} \mathcal{C}_j^u;\theta\right)
    \label{eq:mle}
\end{equation}

\noindent where $\mathcal{C}_j$ represents the NAT conditions for the constrained group $j$, $\mathcal{C}_j^u$ represents the NAT conditions for the unconstrained group $j$, and $\Theta$ denotes all possible global matchings. We assume that the NAT conditions for different constrained groups are mutually independent. Given the NAT conditions for all constrained groups, those for different unconstrained groups are considered mutually conditionally independent. As such, we have

\begin{equation}
\begin{aligned}
    P\left(\bigcap_{j\in \mathcal{G}} \mathcal{C}_j,\bigcap_{j\in \mathcal{G}_u} \mathcal{C}_j^u;\theta\right)
    &=P\left(\bigcap_{j\in \mathcal{G}_u} \mathcal{C}_j^u | \bigcap_{j\in \mathcal{G}} \mathcal{C}_j;\theta\right)P\left(\bigcap_{j\in \mathcal{G}} \mathcal{C}_j;\theta\right) \\
    &=\prod_{j\in \mathcal{G}_u} P\left(\mathcal{C}_j^u|\bigcap_{j\in \mathcal{G}} \mathcal{C}_j;\theta\right) \prod_{j \in \mathcal{G}} P\left(\mathcal{C}_j;\theta\right)
\end{aligned}
\end{equation}

Each product term can be calculated using the results from previous MCMC sampling results and Equation~\ref{eq:mcmc_integral}:

\begin{gather}
    P(\mathcal{C}_j;\theta)\approx V_j \frac{1}{N_j} \sum_{i=1}^{N_j} f(\mathbf{t}_{j,i};\theta),\;j\in \mathcal{G}\\
    P\left(\mathcal{C}_j^u | \bigcap_{j\in \mathcal{G}} \mathcal{C}_j;\theta\right)\approx V_j \frac{1}{N_j} \sum_{i=1}^{N_j} f(\mathbf{t}_{j,i};\theta),\;j\in \mathcal{G}_u
\end{gather}

\noindent where  $\mathbf{t}_{j,i}$ is the $i$-th sampling point for group $j$, with each element $t_k^a$ representing vehicle $k$'s NAT. $V_j$ is the volume of the polytope represented by the NAT conditions of group $j$. $N_j$ is the number of samples of group $j$. Note that for unconstrained groups, changes in the NAT conditions of adjacent constrained groups can cause changes in their own NAT conditions, theoretically requiring resampling. However, for simplicity, this effect is ignored in this context.

Given the global match $\theta$, the upstream departure times $t_k^u$ for all vehicles $k \in K$ in group $j$ are known, and we have

\begin{equation}
\begin{aligned}
    f(\mathbf{t}_{j,i};\theta)
    &=\prod_{k\in K} f_k(t_k^a) \\ 
    &=\prod_{k\in K} g_k(t_k^a - t_k^u)
\end{aligned}
\end{equation}

Therefore, the log-likelihood function corresponding to Equation~\ref{eq:mle} can be expressed as

\begin{equation}
    l(\theta)=
    \sum_{j\in \mathcal{G} \cup \mathcal{G}_u}\log \sum_{i=1}^{N_j} f(\mathbf{t}_{j,i};\theta)
\end{equation}

Since matching $\theta$ is only partially related to group $j$, we have

\begin{equation}
\begin{aligned}
    l(\theta)&=
    \sum_{j\in \mathcal{G} \cup \mathcal{G}_u}\log \sum_{i=1}^{N_j} f(\mathbf{t}_{j,i};\theta_j) \\
    &= \sum_{j\in \mathcal{G} \cup \mathcal{G}_u} l_j(\theta_j)
\end{aligned}
\end{equation}

From this, it is clear that the global match's log-likelihood function $l(\theta)$ can be represented as the sum of the log-likelihood functions of different groups $l_j(\theta_j)$.
\end{appendices}

\bibliographystyle{elsarticle-harv}
\bibliography{article}

\begin{thebibliography}{44}
\expandafter\ifx\csname natexlab\endcsname\relax\def\natexlab#1{#1}\fi
\providecommand{\url}[1]{\texttt{#1}}
\providecommand{\href}[2]{#2}
\providecommand{\path}[1]{#1}
\providecommand{\DOIprefix}{doi:}
\providecommand{\ArXivprefix}{arXiv:}
\providecommand{\URLprefix}{URL: }
\providecommand{\Pubmedprefix}{pmid:}
\providecommand{\doi}[1]{\href{http://dx.doi.org/#1}{\path{#1}}}
\providecommand{\Pubmed}[1]{\href{pmid:#1}{\path{#1}}}
\providecommand{\bibinfo}[2]{#2}
\ifx\xfnm\relax \def\xfnm[#1]{\unskip,\space#1}\fi
%Type = Article
\bibitem[{An et~al.(2021)An, Guo, Hong, Lu and Xia}]{an_lane-based_2021}
\bibinfo{author}{An, C.}, \bibinfo{author}{Guo, X.}, \bibinfo{author}{Hong, R.}, \bibinfo{author}{Lu, Z.}, \bibinfo{author}{Xia, J.}, \bibinfo{year}{2021}.
\newblock \bibinfo{title}{Lane-based traffic arrival pattern estimation using license plate recognition data}.
\newblock \bibinfo{journal}{IEEE Intelligent Transportation Systems Magazine} \bibinfo{volume}{14}, \bibinfo{pages}{133--144}.
%Type = Article
\bibitem[{Ban et~al.(2011)Ban, Hao and Sun}]{ban_real_2011}
\bibinfo{author}{Ban, X.}, \bibinfo{author}{Hao, P.}, \bibinfo{author}{Sun, Z.}, \bibinfo{year}{2011}.
\newblock \bibinfo{title}{Real time queue length estimation for signalized intersections using travel times from mobile sensors}.
\newblock \bibinfo{journal}{Transportation Research Part C: Emerging Technologies} \bibinfo{volume}{19}, \bibinfo{pages}{1133--1156}.
\newblock \DOIprefix\doi{10.1016/j.trc.2011.01.002}.
%Type = Inproceedings
\bibitem[{Berry et~al.(1951)Berry, Belmont et~al.}]{berry_distribution_1951}
\bibinfo{author}{Berry, D.S.}, \bibinfo{author}{Belmont, D.M.}, et~al., \bibinfo{year}{1951}.
\newblock \bibinfo{title}{Distribution of vehicle speeds and travel times}, in: \bibinfo{booktitle}{Proceedings of the Second Berkeley Symposium on Mathematical Statistics and Probability}, \bibinfo{organization}{University of California Press}. pp. \bibinfo{pages}{589--602}.
%Type = Article
\bibitem[{Cao et~al.(2024)Cao, Yuan, Ren, Qi, Li, Deng and Ma}]{cao_tracking_2024}
\bibinfo{author}{Cao, Q.}, \bibinfo{author}{Yuan, J.}, \bibinfo{author}{Ren, G.}, \bibinfo{author}{Qi, Y.}, \bibinfo{author}{Li, D.}, \bibinfo{author}{Deng, Y.}, \bibinfo{author}{Ma, W.}, \bibinfo{year}{2024}.
\newblock \bibinfo{title}{Tracking the source of congestion based on a probabilistic sensor flow assignment model}.
\newblock \bibinfo{journal}{Transportation Research Part C: Emerging Technologies} \bibinfo{volume}{165}, \bibinfo{pages}{104736}.
%Type = Article
\bibitem[{Chen et~al.(2018)Chen, Dwivedi, Wainwright and Yu}]{chen_fast_2018}
\bibinfo{author}{Chen, Y.}, \bibinfo{author}{Dwivedi, R.}, \bibinfo{author}{Wainwright, M.J.}, \bibinfo{author}{Yu, B.}, \bibinfo{year}{2018}.
\newblock \bibinfo{title}{Fast {MCMC} sampling algorithms on polytopes}.
\newblock \bibinfo{journal}{Journal of Machine Learning Research} \bibinfo{volume}{19}, \bibinfo{pages}{1--86}.
%Type = Article
\bibitem[{Cheng et~al.(2012)Cheng, Qin, Jin and Ran}]{cheng_exploratory_2012}
\bibinfo{author}{Cheng, Y.}, \bibinfo{author}{Qin, X.}, \bibinfo{author}{Jin, J.}, \bibinfo{author}{Ran, B.}, \bibinfo{year}{2012}.
\newblock \bibinfo{title}{An exploratory shockwave approach to estimating queue length using probe trajectories}.
\newblock \bibinfo{journal}{Journal of Intelligent Transportation Systems} \bibinfo{volume}{16}, \bibinfo{pages}{12--23}.
\newblock \DOIprefix\doi{10.1080/15472450.2012.639637}.
%Type = Article
\bibitem[{Comert and Cetin(2011)}]{comert_analytical_2011}
\bibinfo{author}{Comert, G.}, \bibinfo{author}{Cetin, M.}, \bibinfo{year}{2011}.
\newblock \bibinfo{title}{Analytical evaluation of the error in queue length estimation at traffic signals from probe vehicle data}.
\newblock \bibinfo{journal}{IEEE Transactions on Intelligent Transportation Systems} \bibinfo{volume}{12}, \bibinfo{pages}{563--573}.
\newblock \DOIprefix\doi{10.1109/TITS.2011.2113375}.
%Type = Article
\bibitem[{Kannan and Narayanan(2012)}]{kannan_random_2012}
\bibinfo{author}{Kannan, R.}, \bibinfo{author}{Narayanan, H.}, \bibinfo{year}{2012}.
\newblock \bibinfo{title}{Random walks on polytopes and an affine interior point method for linear programming}.
\newblock \bibinfo{journal}{Mathematics of Operations Research} \bibinfo{volume}{37}, \bibinfo{pages}{1--20}.
%Type = Article
\bibitem[{Knuth(2000)}]{knuth_dancing_2000}
\bibinfo{author}{Knuth, D.E.}, \bibinfo{year}{2000}.
\newblock \bibinfo{title}{Dancing links}.
\newblock \bibinfo{journal}{arXiv preprint cs/0011047} .
%Type = Article
\bibitem[{Li et~al.(2017)Li, Tang, Yao and Li}]{li_real-time_2017}
\bibinfo{author}{Li, F.}, \bibinfo{author}{Tang, K.}, \bibinfo{author}{Yao, J.}, \bibinfo{author}{Li, K.}, \bibinfo{year}{2017}.
\newblock \bibinfo{title}{Real-time queue length estimation for signalized intersections using vehicle trajectory data}.
\newblock \bibinfo{journal}{Transportation Research Record} \bibinfo{volume}{2623}, \bibinfo{pages}{49--59}.
\newblock \DOIprefix\doi{10.3141/2623-06}.
%Type = Article
\bibitem[{Li et~al.(2023)Li, Tang, Chen and Liu}]{li_traffic_2023}
\bibinfo{author}{Li, M.}, \bibinfo{author}{Tang, J.}, \bibinfo{author}{Chen, Q.}, \bibinfo{author}{Liu, Y.}, \bibinfo{year}{2023}.
\newblock \bibinfo{title}{Traffic arrival pattern estimation at urban intersection using license plate recognition data}.
\newblock \bibinfo{journal}{Physica A: Statistical Mechanics and its Applications} \bibinfo{volume}{625}, \bibinfo{pages}{128995}.
\newblock \DOIprefix\doi{10.1016/j.physa.2023.128995}.
%Type = Article
\bibitem[{Liu et~al.(2009)Liu, Wu, Ma and Hu}]{liu_real-time_2009}
\bibinfo{author}{Liu, H.X.}, \bibinfo{author}{Wu, X.}, \bibinfo{author}{Ma, W.}, \bibinfo{author}{Hu, H.}, \bibinfo{year}{2009}.
\newblock \bibinfo{title}{Real-time queue length estimation for congested signalized intersections}.
\newblock \bibinfo{journal}{Transportation Research Part C: Emerging Technologies} \bibinfo{volume}{17}, \bibinfo{pages}{412--427}.
\newblock \DOIprefix\doi{10.1016/j.trc.2009.02.003}.
%Type = Article
\bibitem[{Luo et~al.(2019)Luo, Ma, Jin, Gong and Wang}]{luo_queue_2019}
\bibinfo{author}{Luo, X.}, \bibinfo{author}{Ma, D.}, \bibinfo{author}{Jin, S.}, \bibinfo{author}{Gong, Y.}, \bibinfo{author}{Wang, D.}, \bibinfo{year}{2019}.
\newblock \bibinfo{title}{Queue length estimation for signalized intersections using license plate recognition data}.
\newblock \bibinfo{journal}{IEEE Intelligent Transportation Systems Magazine} \bibinfo{volume}{11}, \bibinfo{pages}{209--220}.
\newblock \DOIprefix\doi{10.1109/MITS.2019.2919541}.
%Type = Article
\bibitem[{Ma et~al.(2018)Ma, Luo, Jin, Guo and Wang}]{ma_estimating_2018}
\bibinfo{author}{Ma, D.}, \bibinfo{author}{Luo, X.}, \bibinfo{author}{Jin, S.}, \bibinfo{author}{Guo, W.}, \bibinfo{author}{Wang, D.}, \bibinfo{year}{2018}.
\newblock \bibinfo{title}{Estimating maximum queue length for traffic lane groups using travel times from video-imaging data}.
\newblock \bibinfo{journal}{IEEE Intelligent Transportation Systems Magazine} \bibinfo{volume}{10}, \bibinfo{pages}{123--134}.
\newblock \DOIprefix\doi{10.1109/MITS.2018.2842047}.
%Type = Article
\bibitem[{Ma et~al.(2017)Ma, Luo, Li, Jin, Guo and Wang}]{ma_traffic_2017}
\bibinfo{author}{Ma, D.}, \bibinfo{author}{Luo, X.}, \bibinfo{author}{Li, W.}, \bibinfo{author}{Jin, S.}, \bibinfo{author}{Guo, W.}, \bibinfo{author}{Wang, D.}, \bibinfo{year}{2017}.
\newblock \bibinfo{title}{Traffic demand estimation for lane groups at signal-controlled intersections using travel times from video-imaging detectors}.
\newblock \bibinfo{journal}{IET Intelligent Transport Systems} \bibinfo{volume}{11}, \bibinfo{pages}{222--229}.
\newblock \DOIprefix\doi{10.1049/iet-its.2016.0233}.
%Type = Article
\bibitem[{Ma et~al.(2021)Ma, Xiao, Song, Ma and Jin}]{ma_back-pressure-based_2021}
\bibinfo{author}{Ma, D.}, \bibinfo{author}{Xiao, J.}, \bibinfo{author}{Song, X.}, \bibinfo{author}{Ma, X.}, \bibinfo{author}{Jin, S.}, \bibinfo{year}{2021}.
\newblock \bibinfo{title}{A back-pressure-based model with fixed phase sequences for traffic signal optimization under oversaturated networks}.
\newblock \bibinfo{journal}{IEEE Transactions on Intelligent Transportation Systems} \bibinfo{volume}{22}, \bibinfo{pages}{5577--5588}.
\newblock \DOIprefix\doi{10.1109/TITS.2020.2987917}.
%Type = Article
\bibitem[{Mo et~al.(2020)Mo, Li and Dai}]{mo_estimating_2020}
\bibinfo{author}{Mo, B.}, \bibinfo{author}{Li, R.}, \bibinfo{author}{Dai, J.}, \bibinfo{year}{2020}.
\newblock \bibinfo{title}{Estimating dynamic origin–destination demand: {A} hybrid framework using license plate recognition data}.
\newblock \bibinfo{journal}{Computer-Aided Civil and Infrastructure Engineering} \bibinfo{volume}{35}, \bibinfo{pages}{734--752}.
\newblock \DOIprefix\doi{https://doi.org/10.1111/mice.12526}.
%Type = Article
\bibitem[{Mo et~al.(2017)Mo, Li and Zhan}]{mo_speed_2017}
\bibinfo{author}{Mo, B.}, \bibinfo{author}{Li, R.}, \bibinfo{author}{Zhan, X.}, \bibinfo{year}{2017}.
\newblock \bibinfo{title}{Speed profile estimation using license plate recognition data}.
\newblock \bibinfo{journal}{Transportation Research Part C: Emerging Technologies} \bibinfo{volume}{82}, \bibinfo{pages}{358--378}.
\newblock \DOIprefix\doi{10.1016/j.trc.2017.07.006}.
%Type = Article
\bibitem[{Noaeen et~al.(2021)Noaeen, Mohajerpoor, Far and Ramezani}]{noaeen_real-time_2021}
\bibinfo{author}{Noaeen, M.}, \bibinfo{author}{Mohajerpoor, R.}, \bibinfo{author}{Far, B.H.}, \bibinfo{author}{Ramezani, M.}, \bibinfo{year}{2021}.
\newblock \bibinfo{title}{Real-time decentralized traffic signal control for congested urban networks considering queue spillbacks}.
\newblock \bibinfo{journal}{Transportation Research Part C: Emerging Technologies} \bibinfo{volume}{133}.
\newblock \DOIprefix\doi{10.1016/j.trc.2021.103407}.
%Type = Article
\bibitem[{Ramezani and Geroliminis(2015)}]{ramezani_queue_2015}
\bibinfo{author}{Ramezani, M.}, \bibinfo{author}{Geroliminis, N.}, \bibinfo{year}{2015}.
\newblock \bibinfo{title}{Queue profile estimation in congested urban networks with probe data}.
\newblock \bibinfo{journal}{Computer-Aided Civil and Infrastructure Engineering} \bibinfo{volume}{30}, \bibinfo{pages}{414--432}.
\newblock \DOIprefix\doi{10.1111/mice.12095}.
%Type = Article
\bibitem[{Rao et~al.(2018)Rao, Wu, Xia, Ou and Kluger}]{rao_origin-destination_2018}
\bibinfo{author}{Rao, W.}, \bibinfo{author}{Wu, Y.J.}, \bibinfo{author}{Xia, J.}, \bibinfo{author}{Ou, J.}, \bibinfo{author}{Kluger, R.}, \bibinfo{year}{2018}.
\newblock \bibinfo{title}{Origin-destination pattern estimation based on trajectory reconstruction using automatic license plate recognition data}.
\newblock \bibinfo{journal}{Transportation Research Part C: Emerging Technologies} \bibinfo{volume}{95}, \bibinfo{pages}{29--46}.
\newblock \DOIprefix\doi{10.1016/j.trc.2018.07.002}.
%Type = Article
\bibitem[{Sharma et~al.(2007)Sharma, Bullock and Bonneson}]{sharma_inputoutput_2007}
\bibinfo{author}{Sharma, A.}, \bibinfo{author}{Bullock, D.M.}, \bibinfo{author}{Bonneson, J.A.}, \bibinfo{year}{2007}.
\newblock \bibinfo{title}{Input–output and hybrid techniques for real-time prediction of delay and maximum queue length at signalized intersections}.
\newblock \bibinfo{journal}{Transportation Research Record} \bibinfo{volume}{2035}, \bibinfo{pages}{69--80}.
\newblock \DOIprefix\doi{10.3141/2035-08}.
%Type = Article
\bibitem[{Skabardonis and Geroliminis(2008)}]{skabardonis_real-time_2008}
\bibinfo{author}{Skabardonis, A.}, \bibinfo{author}{Geroliminis, N.}, \bibinfo{year}{2008}.
\newblock \bibinfo{title}{Real-time monitoring and control on signalized arterials}.
\newblock \bibinfo{journal}{Journal of Intelligent Transportation Systems} \bibinfo{volume}{12}, \bibinfo{pages}{64--74}.
\newblock \DOIprefix\doi{10.1080/15472450802023337}.
%Type = Article
\bibitem[{Tan et~al.(2024a)Tan, Cao, Ban and Tang}]{tan2024connected}
\bibinfo{author}{Tan, C.}, \bibinfo{author}{Cao, Y.}, \bibinfo{author}{Ban, X.}, \bibinfo{author}{Tang, K.}, \bibinfo{year}{2024}a.
\newblock \bibinfo{title}{Connected vehicle data-driven fixed-time traffic signal control considering cyclic time-dependent vehicle arrivals based on cumulative flow diagram}.
\newblock \bibinfo{journal}{IEEE Transactions on Intelligent Transportation Systems} .
%Type = Article
\bibitem[{Tan et~al.(2024b)Tan, Ding, Yang, Zhu and Tang}]{tan_connected_2024}
\bibinfo{author}{Tan, C.}, \bibinfo{author}{Ding, Y.}, \bibinfo{author}{Yang, K.}, \bibinfo{author}{Zhu, H.}, \bibinfo{author}{Tang, K.}, \bibinfo{year}{2024}b.
\newblock \bibinfo{title}{Connected vehicle data-driven robust optimization for traffic signal timing: Modeling traffic flow variability and errors}.
\newblock \bibinfo{journal}{arXiv preprint arXiv:2406.14108} .
%Type = Article
\bibitem[{Tan et~al.(2020)Tan, Liu, Wu, Cao and Tang}]{tan_fuzing_2020}
\bibinfo{author}{Tan, C.}, \bibinfo{author}{Liu, L.}, \bibinfo{author}{Wu, H.}, \bibinfo{author}{Cao, Y.}, \bibinfo{author}{Tang, K.}, \bibinfo{year}{2020}.
\newblock \bibinfo{title}{Fuzing license plate recognition data and vehicle trajectory data for lane-based queue length estimation at signalized intersections}.
\newblock \bibinfo{journal}{Journal of Intelligent Transportation Systems} \bibinfo{volume}{24}, \bibinfo{pages}{449--466}.
\newblock \DOIprefix\doi{10.1080/15472450.2020.1732217}.
%Type = Article
\bibitem[{Tan et~al.(2022a)Tan, Wu, Tang and Tan}]{tan_extendable_2022}
\bibinfo{author}{Tan, C.}, \bibinfo{author}{Wu, H.}, \bibinfo{author}{Tang, K.}, \bibinfo{author}{Tan, C.}, \bibinfo{year}{2022}a.
\newblock \bibinfo{title}{An extendable gaussian mixture model for lane-based queue length estimation based on license plate recognition data}.
\newblock \bibinfo{journal}{Journal of Advanced Transportation} \bibinfo{volume}{2022}, \bibinfo{pages}{e5119209}.
\newblock \DOIprefix\doi{10.1155/2022/5119209}.
%Type = Article
\bibitem[{Tan and Yang(2024)}]{tan2024privacy}
\bibinfo{author}{Tan, C.}, \bibinfo{author}{Yang, K.}, \bibinfo{year}{2024}.
\newblock \bibinfo{title}{Privacy-preserving adaptive traffic signal control in a connected vehicle environment}.
\newblock \bibinfo{journal}{Transportation research part C: emerging technologies} \bibinfo{volume}{158}, \bibinfo{pages}{104453}.
%Type = Article
\bibitem[{Tan et~al.(2022b)Tan, Yao, Ban and Tang}]{tan_cumulative_2022}
\bibinfo{author}{Tan, C.}, \bibinfo{author}{Yao, J.}, \bibinfo{author}{Ban, X.}, \bibinfo{author}{Tang, K.}, \bibinfo{year}{2022}b.
\newblock \bibinfo{title}{Cumulative flow diagram estimation and prediction based on sampled vehicle trajectories at signalized intersections}.
\newblock \bibinfo{journal}{IEEE Transactions on Intelligent Transportation Systems} \bibinfo{volume}{23}, \bibinfo{pages}{11325--11337}.
\newblock \DOIprefix\doi{10.1109/TITS.2021.3102750}.
%Type = Article
\bibitem[{Tan et~al.(2021)Tan, Yao, Tang and Sun}]{tan_cycle-based_2021}
\bibinfo{author}{Tan, C.}, \bibinfo{author}{Yao, J.}, \bibinfo{author}{Tang, K.}, \bibinfo{author}{Sun, J.}, \bibinfo{year}{2021}.
\newblock \bibinfo{title}{Cycle-based queue length estimation for signalized intersections using sparse vehicle trajectory data}.
\newblock \bibinfo{journal}{IEEE Transactions on Intelligent Transportation Systems} \bibinfo{volume}{22}, \bibinfo{pages}{91--106}.
\newblock \DOIprefix\doi{10.1109/TITS.2019.2954937}.
%Type = Article
\bibitem[{Tang et~al.(2021)Tang, Cao, Chen, Yao, Tan and Sun}]{tang_dynamic_2021}
\bibinfo{author}{Tang, K.}, \bibinfo{author}{Cao, Y.}, \bibinfo{author}{Chen, C.}, \bibinfo{author}{Yao, J.}, \bibinfo{author}{Tan, C.}, \bibinfo{author}{Sun, J.}, \bibinfo{year}{2021}.
\newblock \bibinfo{title}{Dynamic origin-destination flow estimation using automatic vehicle identification data: {A} {3D} convolutional neural network approach}.
\newblock \bibinfo{journal}{Computer-Aided Civil and Infrastructure Engineering} \bibinfo{volume}{36}, \bibinfo{pages}{30--46}.
\newblock \DOIprefix\doi{10.1111/mice.12559}.
%Type = Article
\bibitem[{Tang et~al.(2022)Tang, Wu, Yao, Tan and Ji}]{tang_lane-based_2022}
\bibinfo{author}{Tang, K.}, \bibinfo{author}{Wu, H.}, \bibinfo{author}{Yao, J.}, \bibinfo{author}{Tan, C.}, \bibinfo{author}{Ji, Y.}, \bibinfo{year}{2022}.
\newblock \bibinfo{title}{Lane-based queue length estimation at signalized intersections using single-section license plate recognition data}.
\newblock \bibinfo{journal}{Transportmetrica B: Transport Dynamics} \bibinfo{volume}{10}, \bibinfo{pages}{293--311}.
\newblock \DOIprefix\doi{10.1080/21680566.2021.1991504}.
%Type = Article
\bibitem[{Telgen(1983)}]{telgen_identifying_1983}
\bibinfo{author}{Telgen, J.}, \bibinfo{year}{1983}.
\newblock \bibinfo{title}{Identifying redundant constraints and implicit equalities in systems of linear constraints}.
\newblock \bibinfo{journal}{Management Science} \bibinfo{volume}{29}, \bibinfo{pages}{1209--1222}.
\newblock \DOIprefix\doi{10.1287/mnsc.29.10.1209}.
%Type = Inproceedings
\bibitem[{Vaidya(1989)}]{vaidya_new_1989}
\bibinfo{author}{Vaidya, P.M.}, \bibinfo{year}{1989}.
\newblock \bibinfo{title}{A new algorithm for minimizing convex functions over convex sets}, in: \bibinfo{booktitle}{30th {Annual} {Symposium} on {Foundations} of {Computer} {Science}}, pp. \bibinfo{pages}{338--343}.
\newblock \DOIprefix\doi{10.1109/SFCS.1989.63500}.
%Type = Article
\bibitem[{Vigos et~al.(2008)Vigos, Papageorgiou and Wang}]{vigos_real-time_2008}
\bibinfo{author}{Vigos, G.}, \bibinfo{author}{Papageorgiou, M.}, \bibinfo{author}{Wang, Y.}, \bibinfo{year}{2008}.
\newblock \bibinfo{title}{Real-time estimation of vehicle-count within signalized links}.
\newblock \bibinfo{journal}{Transportation Research Part C: Emerging Technologies} \bibinfo{volume}{16}, \bibinfo{pages}{18--35}.
\newblock \DOIprefix\doi{10.1016/j.trc.2007.06.002}.
%Type = Article
\bibitem[{Wu et~al.(2024)Wu, Luo, Oguchi, Tang and Zhu}]{wu_stochastic_2024}
\bibinfo{author}{Wu, H.}, \bibinfo{author}{Luo, L.}, \bibinfo{author}{Oguchi, T.}, \bibinfo{author}{Tang, K.}, \bibinfo{author}{Zhu, H.}, \bibinfo{year}{2024}.
\newblock \bibinfo{title}{Stochastic queue profile estimation using license plate recognition data}.
\newblock \bibinfo{journal}{Physica A: Statistical Mechanics and its Applications} \bibinfo{volume}{643}, \bibinfo{pages}{129790}.
\newblock \DOIprefix\doi{10.1016/j.physa.2024.129790}.
%Type = Inproceedings
\bibitem[{Wu et~al.(2019)Wu, Yao, Liu, Cao and Tang}]{wu_left-turn_2019}
\bibinfo{author}{Wu, H.}, \bibinfo{author}{Yao, J.}, \bibinfo{author}{Liu, L.}, \bibinfo{author}{Cao, Y.}, \bibinfo{author}{Tang, K.}, \bibinfo{year}{2019}.
\newblock \bibinfo{title}{Left-turn spillback identification based on license plate recognition data}, in: \bibinfo{booktitle}{Proceedings of the 98th {Annual} {Meeting} {Transportation} {Research} {Board}}, \bibinfo{publisher}{Transportation Research Board}, \bibinfo{address}{Washington, DC}.
%Type = Article
\bibitem[{Wu et~al.(2010)Wu, Liu and Gettman}]{wu_identification_2010}
\bibinfo{author}{Wu, X.}, \bibinfo{author}{Liu, H.X.}, \bibinfo{author}{Gettman, D.}, \bibinfo{year}{2010}.
\newblock \bibinfo{title}{Identification of oversaturated intersections using high-resolution traffic signal data}.
\newblock \bibinfo{journal}{Transportation Research Part C: Emerging Technologies} \bibinfo{volume}{18}, \bibinfo{pages}{626--638}.
\newblock \DOIprefix\doi{10.1016/j.trc.2010.01.003}.
%Type = Article
\bibitem[{Yang and Sun(2015)}]{yang_vehicle_2015}
\bibinfo{author}{Yang, J.}, \bibinfo{author}{Sun, J.}, \bibinfo{year}{2015}.
\newblock \bibinfo{title}{Vehicle path reconstruction using automatic vehicle identification data: {An} integrated particle filter and path flow estimator}.
\newblock \bibinfo{journal}{Transportation Research Part C: Emerging Technologies} \bibinfo{volume}{58}, \bibinfo{pages}{107--126}.
\newblock \DOIprefix\doi{10.1016/j.trc.2015.07.003}.
%Type = Article
\bibitem[{Yao et~al.(2020)Yao, Jiang, Zhao, Luo and Peng}]{yao_dynamic_2020}
\bibinfo{author}{Yao, Z.}, \bibinfo{author}{Jiang, Y.}, \bibinfo{author}{Zhao, B.}, \bibinfo{author}{Luo, X.}, \bibinfo{author}{Peng, B.}, \bibinfo{year}{2020}.
\newblock \bibinfo{title}{A dynamic optimization method for adaptive signal control in a connected vehicle environment}.
\newblock \bibinfo{journal}{Journal of Intelligent Transportation Systems} \bibinfo{volume}{24}, \bibinfo{pages}{184--200}.
\newblock \DOIprefix\doi{10.1080/15472450.2019.1643723}.
%Type = Article
\bibitem[{Yin et~al.(2021)Yin, Chen, Tang and Sun}]{yin_queue_2021}
\bibinfo{author}{Yin, J.}, \bibinfo{author}{Chen, P.}, \bibinfo{author}{Tang, K.}, \bibinfo{author}{Sun, J.}, \bibinfo{year}{2021}.
\newblock \bibinfo{title}{Queue intensity adaptive signal control for isolated intersection based on vehicle trajectory data}.
\newblock \bibinfo{journal}{Journal of Advanced Transportation} \bibinfo{volume}{2021}, \bibinfo{pages}{8838922}.
\newblock \DOIprefix\doi{10.1155/2021/8838922}.
%Type = Article
\bibitem[{Zhan et~al.(2015)Zhan, Li and Ukkusuri}]{zhan_lane-based_2015}
\bibinfo{author}{Zhan, X.}, \bibinfo{author}{Li, R.}, \bibinfo{author}{Ukkusuri, S.V.}, \bibinfo{year}{2015}.
\newblock \bibinfo{title}{Lane-based real-time queue length estimation using license plate recognition data}.
\newblock \bibinfo{journal}{Transportation Research Part C: Emerging Technologies} \bibinfo{volume}{57}, \bibinfo{pages}{85--102}.
\newblock \DOIprefix\doi{10.1016/j.trc.2015.06.001}.
%Type = Article
\bibitem[{Zhan et~al.(2020)Zhan, Li and Ukkusuri}]{zhan_link-based_2020}
\bibinfo{author}{Zhan, X.}, \bibinfo{author}{Li, R.}, \bibinfo{author}{Ukkusuri, S.V.}, \bibinfo{year}{2020}.
\newblock \bibinfo{title}{Link-based traffic state estimation and prediction for arterial networks using license-plate recognition data}.
\newblock \bibinfo{journal}{Transportation Research Part C: Emerging Technologies} \bibinfo{volume}{117}, \bibinfo{pages}{102660}.
\newblock \DOIprefix\doi{10.1016/j.trc.2020.102660}.
%Type = Article
\bibitem[{Zhang et~al.(2020)Zhang, Liu, Chen, Yu and Wang}]{zhang_cycle-based_2020}
\bibinfo{author}{Zhang, H.}, \bibinfo{author}{Liu, H.X.}, \bibinfo{author}{Chen, P.}, \bibinfo{author}{Yu, G.}, \bibinfo{author}{Wang, Y.}, \bibinfo{year}{2020}.
\newblock \bibinfo{title}{Cycle-based end of queue estimation at signalized intersections using low-penetration-rate vehicle trajectories}.
\newblock \bibinfo{journal}{IEEE Transactions on Intelligent Transportation Systems} \bibinfo{volume}{21}, \bibinfo{pages}{3257--3272}.
\newblock \DOIprefix\doi{10.1109/TITS.2019.2925111}.

\end{thebibliography}
\end{document}